\documentclass{article}
\pdfoutput=1
\usepackage{amsmath}
\usepackage{amssymb}
\usepackage{graphicx}
\usepackage{color}
\usepackage[usenames,dvipsnames,svgnames,table]{xcolor}
\usepackage{soul}
\usepackage{subfig}
\usepackage{ulem}

\usepackage{booktabs}
\usepackage{multirow}
\usepackage{makecell}

\usepackage[section]{placeins}

\pdfpageheight\paperheight
\pdfpagewidth\paperwidth

\usepackage{jcappub}

\definecolor{darkred}{RGB}{175,0,0}

\newcommand{\beq}{\begin{equation}}

\newcommand{\eeq}{\end{equation}}

\usepackage{array}
\def\ga{ \mathrel{\rlap{\raise 0.59ex
 \hbox{$>$}}{\lower 0.59ex \hbox{$\sim$}}}}

\newcommand{\ra}[1]{\renewcommand{\arraystretch}{#1}}

\begin{document}
\title{Enlightening the dark ages with dark matter}
\author[a,b]{Katie Short}
\author[c,a,b]{Jos\' e Luis Bernal}
\author[d,a]{Alvise Raccanelli}
\author[a,e]{Licia Verde}
\author[f]{Jens Chluba}

\affiliation[a]{ICC, University of Barcelona, IEEC-UB, Mart\' i i Franqu\` es, 1, E08028
Barcelona, Spain}
\affiliation[b]{Dept. de F\' isica Qu\` antica i Astrof\' isica, Universitat de Barcelona, Mart\' i i Franqu\` es 1, E08028 Barcelona,
Spain}
\affiliation[c]{Department of Physics and Astronomy, Johns Hopkins University, 3400 North Charles Street, Baltimore, Maryland 21218, USA}
\affiliation[d]{Theoretical Physics Department, CERN, 1 Esplanade des Particules, CH-1211 Geneva 23, Switzerland}
\affiliation[f]{ICREA, Pg. Llu\' is Companys 23, 08010 Barcelona, Spain}
\affiliation[e]{Jodrell Bank Centre for Astrophysics, School of Physics and Astronomy, The University of Manchester, Manchester M13 9PL, U.K.}

\emailAdd{katie.short@icc.ub.edu}
\emailAdd{jbernal2@jhu.edu}
\emailAdd{alvise.raccanelli@cern.ch}
\emailAdd{liciaverde@icc.ub.edu}

\abstract{Constraints on dark matter annihilation or decay offer unique insights into the nature of dark matter. We illustrate how surveys dedicated to detect the highly redshifted 21 cm signal from the dark ages will offer a new window into properties of particle dark matter. The 21 cm intensity mapping signal and its fluctuations are sensitive to energy injection from annihilation or decay of long-lived particles in a way that is complementary to other probes. We present forecasted constraints from forthcoming and next-generation radio surveys. We show that, while SKA might be capable of a detection for some cases, the most promising opportunity to detect dark matter in the 21 cm intensity mapping signal is with a futuristic radio array on the lunar far-side, with the potential to detect a signal many orders of magnitude weaker than current or maximal constraints from other probes.
}
\maketitle

\hypersetup{pageanchor=true}

\section{Introduction}\label{sec:Introduction}

Understanding the nature of the Dark Matter (DM) is one of the primary goals of research in cosmology and astro-particle physics. DM is required to explain a wide variety of both cosmological and astrophysical observations, such as the Cosmic Microwave Background (CMB) and the growth of Large Scale Structure (LSS), to galaxy rotation curves and the lensing of galaxies and clusters~(\cite{Bergstrom:2000pn, Bertone_2005} and references therein). However, despite relentless theoretical and observational efforts, the nature of this form of matter remains unknown. There is a vast array of DM candidates, many with electromagnetic signatures either through coupling to Standard Model particles via decay, annihilation and scattering, or due to standard astrophysical processes such as primordial black holes~\cite{Ali-Haimoud:2019khd, Gluscevic:2019yal}. While DM must be predominantly stable over cosmological timescales, it is possible to have particle DM models with a very long yet finite decay lifetime (e.g., sterile neutrinos~\cite{Abazajian:2012ys,Boyarsky:2018tvu}, axinos~\cite{KIM200218, Hamaguchi:2017ihw}, R-parity violating SUSY models~\cite{BEREZINSKY1991382}) or where a small fraction of the total DM is decaying (e.g., 
 excited metastable species~\cite{Finkbeiner:2007kk, Finkbeiner:2014sja}, atomic DM~\cite{Kaplan:2009de}, or some other richer dark sector physics). If the DM consists of a massive particle with weak-scale interactions (thermal relic WIMP~\cite{Jungman:1995df}), then we expect coupling to Standard Model particles via DM self-annihilations.
\par
DM candidates with phenomenology beyond the standard cold DM properties are constrained by both astrophysical and cosmological observations~\cite{Gaskins_2016}, as well as direct detection~\cite{Schumann:2019eaa, Lin:2019uvt} and collider~\cite{Boveia:2018yeb} laboratory experiments. Signatures of particle DM in the low-redshift Universe, such as high energy cosmic rays or \(\gamma\)--rays from annihilation or decay of DM in astrophysical sources within our local environment (dwarf galaxies, Galactic centre), have been searched for and used to place stringent bounds on the particle properties (see e.g.,~\cite{Slatyer:2017sev} and references therein). Non-gravitational effects of particle DM have also been constrained from early Universe observations, such as the CMB power spectrum~\cite{Galli:2009zc, Slatyer:2009yq, Huetsi:2009ex, Galli:2011rz, Finkbeiner:2011dx, Ade:2015xua, Slatyer:2015jla, Poulin:2016anj, Aghanim:2018eyx} and spectral distortions~\cite{Ellis:1984eq, Sarkar:1984tt, Hu:1993gc,McDonald:2000bk, Chluba:2011hw, Chluba:2013wsa, Chluba:2013pya, Poulin:2016anj}, as well as Big-Bang Nucleosynthesis~\cite{Iocco:2008va, Pospelov:2010hj, Poulin:2015opa}.
\par
Probes such as Big-Bang Nucleosynthesis and CMB spectral distortions dominate constraints for DM models with short decay lifetimes (\(\tau \lesssim 10^{12} \, \rm{s}\)), while the CMB angular power spectrum is sensitive to longer lifetimes and strongly disfavours lifetimes \(\tau \lesssim 10^{24}-10^{25} \, \mathrm{s}\) if all DM decays~\cite{Poulin:2016anj, Slatyer:2016qyl, Lucca:2019rxf}. More stringent limits have been set on the decay lifetime in the GeV--TeV mass regime via astrophysical probes; the strongest constraints come from the lack of signal in the Isotropic Gamma-Ray Background above known astrophysical sources~\cite{Cohen:2016uyg, Blanco:2018esa}. However, these constraints are commonly affected by large uncertainties in the diffuse Galactic background modelling, the Milky Way's DM halo profile, and features of Galactic cosmic-ray propagation. Constraints are typically weaker for light (below 1 GeV) decaying DM. Similarly, measurements of CMB anisotropies have been used to place constraints on the amount of energy injection due to annihilation of DM particles~\cite{Galli:2009zc, Giesen:2012rp, Lopez-Honorez:2013lcm, Ade:2015xua, Slatyer:2015jla, Kawasaki:2015peu,Aghanim:2018eyx}. As in the case of DM decay, indirect detection constraints can be stronger for super-GeV (\(>\)1 GeV) annihilating DM than those found from CMB measurements (e.g.~\cite{Abramowski:2014tra, Ackermann:2015zua, Ahnen:2016qkx, Fermi-LAT:2016uux, Archambault:2017wyh}), but are generally model-dependent and less robust due to the associated astrophysical uncertainties. Despite these constraints, there remains a large portion of viable parameter space for DM decays or annihilations to the visible sector left unexplored~\cite{Leane:2018kjk}.
\par
Future searches must aim to beat and improve these limits, and eventually achieve detection to characterize the nature of DM.  However, even with drastically more sensitive future CMB anisotropy instruments, the forecasted constraints on DM annihilation are not expected to drastically improve~\cite{Madhavacheril:2013cna,Green:2018pmd}. It is thus vital that new probes are explored. The advent of line-intensity mapping (LIM) opens up a promising new path for the study of both cosmology and astrophysics~\cite{Kovetz:2017agg, Kovetz:2019uss}. Appropriate analyses of LIM observations will allow to probe cosmology beyond the reach of galaxy surveys (see e.g.,~\cite{Bernal:2019jdo}), bridging the unexplored epoch of the Universe between redshift $z\sim 3-5$ and the CMB (see e.g.,~\cite{Bernal:2019gfq,Munoz:2019fkt}). The 21 cm line~\cite{Furlanetto:2006jb, Pritchard:2011xb} offers a direct way to survey the cosmic dark ages, spanning the evolution of the Universe following the end of recombination until the formation of the first luminous objects, with an expected signal in principle reaching up to $z\sim 500$~\cite{Breysse:2018slj}. Electromagnetic energy injection induced by DM decay or annihilation during the dark ages would alter the thermal and ionization history of the intergalactic medium (IGM) and leave a potentially detectable imprint in the 21 cm LIM signal and its fluctuations (as discussed in~\cite{Furlanetto:2006wp, Shchekinov:2006eb, Valdes:2007cu, Valdes:2012zv,Ali-Haimoud:2013hpa,Evoli:2014pva, Lopez-Honorez:2016sur, Poulin:2016nat}). Since the 21 cm LIM signal is sensitive to the late-time behaviour of the Universe and depends on the \textit{thermal} history, not only the ionization fraction (like CMB), it may provide a powerful complementary cosmological probe of DM with very long lifetimes or low annihilation rates that otherwise evade constraints from the CMB. Furthermore, in contrast to CMB measurements, the onset of structure formation can have an impact on the DM annihilation rate~\cite{Huetsi:2009ex, Cirelli:2009bb, Liu:2016cnk}, boosting the effect in the 21 cm signal. Both the global 21 cm signal during reionization and the cosmic dawn and the power spectrum of its fluctuations have been proposed as a probe for exotic DM (see e.g.,~\cite{Valdes:2012zv, Evoli:2014pva, Poulin:2016anj}). However,  isolating an unambiguous DM signal in this era, given the uncertainties in the physics of reionization, appears challenging~\cite{Cohen:2016jbh, Cohen:2017xpx}. 
\par
On the other hand, targeting the 21 cm signal from the dark ages, although observationally challenging, would circumvent the astrophysical uncertainties related with the cosmic dawn and reionization~\cite{Burns:2019zia, Furlanetto:2019jso}. Furthermore, matter perturbations remain in the linear regime throughout most of this era~\cite{Ali-Haimoud:2013hpa}, which simplifies the modelling of the relevant physics, without complications due to galaxy bias and non-linearities in structure formation. Even when departing from the linear regime at \(z>30\), corrections remain perturbative~\cite{Ali-Haimoud:2013hpa}. Additionally, the 21 cm temperature fluctuations can be observed at many independent redshift slices, the information from which can be combined to perform a tomographic analysis. This kind of observation would thus provide an uncontaminated window to constrain physics beyond the standard model of cosmology (e.g. inflation~\cite{Munoz:2016owz, Pourtsidou:2016ctq, Sekiguchi:2017cdy}, non-gaussianity~\cite{Munoz:2015eqa}), and particularly those involving exotic energy injection such as DM~\cite{Valdes:2012zv, Evoli:2014pva, Poulin:2016anj} and PBH \cite{Bernal:2017nec, Mena:2019nhm}.
\par
The forthcoming Square Kilometre Array\footnote{https://www.skatelescope.org} (SKA)~\cite{Bacon:2018dui} will have the capability to measure the 21 cm LIM signal up to redshifts of \(z\sim 30\). Beyond \(z\gtrsim 30\), the 21 cm signal is redshifted to frequencies that the Earth's atmosphere is opaque to and the signal suffers contamination from terrestrial radio interference, rendering detection impossible unless an extremely precise modelling of these contaminants is achieved. As such, the concept of a low-frequency radio array on the lunar far-side is being developed~\cite{Carilli:2007eb, Jester_2009, Burns:2012bv, LRA_whitepaper} in order to study the 21 cm signal up to high enough redshifts to adequately probe the dark ages and take full advantage of the power of 21 cm LIM tomography. A radio telescope on lunar far-side would be shielded from Earth's ionosphere and terrestrial radio interference. However, as we go to higher redshift (lower frequencies), the radio foregrounds become rapidly brighter (reaching several orders of magnitude larger than the expected signal) and will need to be removed with high precision in order to disentangle the cosmological signal in this frequency range from other sources~\cite{Furlanetto:2006jb}.
\par
In this work, we consider DM particles with decay lifetimes or annihilation cross-sections that just evade current CMB limits. We model the energy injection in a phenomenological way, without modeling all particle physics properties such as the mass and decay or annihilation channels in detail. The purpose of this work is to evaluate the 21 cm LIM power spectrum from the dark ages as an efficient probe of exotic DM models, estimate its potential, and compare it with existing complementary probes such as the CMB. Therefore, we do not aim for a detailed and comprehensive analysis of the DM contribution depending on the specific underlying particle physics, and leave this study for future work.
\par
We begin in Section \ref{sec:21 cm} by reviewing the standard physics of the global 21 cm signal and the angular power spectrum of its fluctuations in the dark ages. In Section \ref{sec:DMdecayannh}, we introduce the effect of DM decay and annihilation in the physics of the dark ages; we review the thermal and ionization history of the IGM, including the effect of exotic energy injection, and show how this leads to changes in the global sky-averaged 21 cm signal and its angular power spectrum. Section \ref{sec:methods} covers the forecasting methodology and the instrumental set-up. Finally, in Section \ref{sec:results}, we present the forecasted detectability of DM decay and annihilation with the forthcoming SKA and different realisations of next-generation radio surveys: a more sensitive ``advanced SKA" (aSKA) and a futuristic ``Lunar Radio Array" (LRA).

\section{Standard 21 cm line-intensity mapping in the dark ages}
\label{sec:21 cm}
In this section we review the standard physics of the 21 cm signal and set out our notation. Reader experts in this subjects can go directly to Section \ref{sec:DMdecayannh}.
\subsection{Global 21 cm signal}
The 21 cm line from neutral hydrogen is triggered by the spin-flip transition between the singlet state and triplet state of the hyperfine structure of the $1s$ ground state. The excitation temperature of this transition is called the \textit{spin temperature} \(T_{s}\), defined via the ratio of the populations
in the two hyperfine levels \(n_{1}/n_{0} = 3 e^{-\frac{T_{\star}}{T_{s}}}\) with energy splitting \(T_{\star} = 0.068 \, \mathrm{K}\). The spin temperature is driven by: (i) absorption/emission due to Compton scattering with ambient CMB photons, (ii) atomic collisions, relevant at high redshift when the IGM is dense, and (iii) resonant scattering with Lyman-\(\alpha\)  photons (the Wouthuysen-Field effect~\cite{Wouthuysen_1952,Field_1959,Hirata:2005mz}). Then, the evolution of the spin temperature follows~\cite{Field_1958}:
\begin{equation}
T_{s} =  \frac{T_{\star} + T_{\mathrm{CMB}}(z) + y_{k}T_{k}(z) + y_{\alpha}T_{\alpha}}{1+y_{k}+y_{\alpha}}, 
\end{equation}
where \(T_{\mathrm{CMB}}\) is the CMB temperature, \(T_{k}\) is the mean kinetic temperature of the cosmic gas,  \(T_{\alpha}\) the color temperature, and \(y_{k}\) and \(y_{\alpha}\) are the kinetic and Lyman-\(\alpha\) coupling terms respectively. The kinetic coupling efficiency $y_k$ is  
\begin{equation}
y_k = \frac{T_\star}{A_{10}T_{k}}(C_H + C_e + C_p),
\end{equation}
with de-excitation rates \(C_i\) (of the triplet) due to collisions with neutral hydrogen, free electrons and free protons, respectively. To determine these rates, we adopt the fitting formulas found in~\cite{Kuhlen:2005cm}:
\begin{equation}
\begin{split}
&C_{H} = n_{H}x_{\rm{HI}}\kappa, \qquad  \qquad
C_{e} =  n_H (1-x_{\rm{HI}}) \gamma_e,
\\
& C_{p} = 3.2n_{H}(1-x_{\rm{HI}}) \kappa ,
\end{split}
\end{equation} 
where $n_H$ is the comoving number density of hydrogen nuclei and $x_{\rm{HI}}$ is the fraction of neutral hydrogen, both in units of $\mathrm{cm^{-3}}$, $\kappa = 3.1 \times 10^{-11}T_{k}^{0.357} \exp\left(\frac{-32}{T_{k}}\right)\,\mathrm{cm^{3} s^{-1} } $ and
\( \log_{10}(\gamma_e/\mathrm{cm^3 s^{-1}}) = -9.607 + 0.5\log_{10}(T_k) \exp[-(\log_{10} T_k)^{4.5}/1800]\) for \(T_k \leq 10^4 \mathrm{K}\) and otherwise, \( \gamma_e (T_k > 10^4 \mathrm{K}) = \gamma_{e} (T_k = 10^4 \mathrm{K})\)~\cite{Liszt:2001kh}. 

Since we focus on the dark ages, before the first stars form, we can safely neglect the Wouthuysen-Field effect and set $y_\alpha=0$. We take the beginning of the dark ages at $z = 30$; nonetheless, in some scenarios star formation might start at higher redshifts, hence introducing higher uncertainties in the measurements at $z\sim 30$ due to its effect on the 21 cm signal (see e.g.,~\cite{Cohen:2016jbh,Cohen:2017xpx}). We implement a simple stellar reionization modeling~\cite{Poulin:2015pna} (whose details can be found in Appendix \ref{sec:reio}), and find that it has no significant impact in our results.
\par
The observed differential brightness temperature, at some frequency \(\nu\), 
 is given by \cite{Field_1959,Furlanetto:2006jb, Pritchard:2011xb},
\begin{multline}
\label{eq:T21}
T_{21}^{\mathrm{obs}} = \frac{T_{s}(z) - T_{\mathrm{CMB}}(z)}{1+z}(1 - e^{-\tau_{\nu_{0}}}) \approx \\
\approx (27\mathrm{mK})(1+\delta_b)x_{\rm{HI}} \left( 1- \frac{T_{\mathrm{CMB}}}{T_{s}} \right) \left( \frac{\Omega_{b}h^{2}}{0.023} \right) \times \left(\frac{1+z}{10} \frac{0.15}{\Omega_{m}h^2}\right)^{1/2} \frac{1}{1 + (1+z) \frac{\partial_{r}v_{r}}{H(z)}},
\end{multline}
where \(\tau_{\nu_{0}}\) is the optical depth of the IGM for the hyperfine transition frequency \(\nu_{0}=1420.4\) MHz,
 \(x_{\rm{HI}}\) is the fraction of neutral hydrogen, \(\delta_b\) is the fractional baryon overdensity, \(\Omega_b\) and \(\Omega_m\) refer to the baryon and matter density parameters, respectively, $H(z)$ is the Hubble parameter, \(h=H_{0}/100\) is the reduced Hubble constant, and \(\partial_{r}v_{r}\) is the comoving (peculiar) velocity gradient along the line of sight. 
 Thus, when \(T_s < T_{\mathrm{CMB}}\), we observe the 21 cm signal in absorption against the CMB spectrum, and in emission for \(T_s > T_{\mathrm{CMB}}\). Beyond \(z \gtrsim 200-500\), collisional coupling is so effective that \(T_{s} = T_{k}\),  while the residual free electron fraction couples the gas to CMB photons via Compton scattering (\(T_{k} = T_{\mathrm{CMB}}\)), and hence $T_{21}^{\rm obs}\sim 0$. With this in mind, we consider as our science case the interval $30\lesssim z\lesssim 200$.

\subsection{Angular power spectrum of the 21 cm fluctuations}
Fluctuations in the \(T_{21}\) signal are expected to be sourced by perturbations in the density and velocity divergence of the hydrogen clouds (which in turn cause fluctuations in the optical depth and spin temperature). For the purposes of this study, in which we consider the dark ages $z \gtrsim 30$, it suffices to treat the perturbations as linear. Additionally, since we are primarily interested in a signal-to-noise calculation, we neglect fluctuations in other quantities (e.g., ionization fraction, gas temperature) which have a sub-dominant effect (see e.g.,~\cite{Pillepich:2006fj} for a full account of these effects up to second-order).
\par 
Here we follow the formalism of~\cite{Ali-Haimoud:2013hpa}, where the fluctuations in the 21 cm differential brightness temperature signal, to linear order, are given by:
\begin{equation}
\label{eq:dT21}
\delta T_{21}(\vec{x}) = \alpha(z) \delta_{b}(\vec{x}) + \bar{T}_{21}(z) \delta_{v}(\vec{x}),
\end{equation}
where \(\alpha(z)=\frac{\mathrm{d}T_{21}}{\mathrm{d}\delta_{b}}\),
\(\delta_{v} \equiv -(1+z) \delta_{r}v_{r}/H(z)\), and we have now dropped the subscript ``obs" from Eq.~\ref{eq:T21}. 
Under the assumption of matter domination, and that the baryons follow perfectly the DM distribution, then \(\delta_b \propto 1/(1+z)\).  Thus, in Fourier space one can write \(\delta_{v}(\bold{k}, z) = ( \hat{k} \cdot \hat{n} )^{2}\delta_{b}(\bold{k},z)\) as defined in Ref.~\cite{Ali-Haimoud:2013hpa}. 
Therefore the transfer function of \(\delta T_{21}\) can be defined as: 
\begin{equation}
\label{eq:Tk}
\mathcal{T}_{\ell}(k,\nu) = \int_0^{\infty} dx \, W_{\nu}(x) [\overline{T}_{21}(z)J_{\ell}(kx) + \alpha(z) j_{\ell}(kx)],
\end{equation}
where \(j_{\ell}(kx)\) are the spherical Bessel functions, and \(J_{\ell}(kx) \equiv - \partial^2 j_{\ell}(kx)\ (\partial kx)^2\) can be written in terms of spherical Bessel functions\footnote{$ J_{\ell} = \frac{-\ell (\ell -1)}{4\ell^2 -1} j_{\ell -2}(kx) + \frac{2\ell^2 + 2\ell -1}{4\ell^2 + 4\ell - 3} j_{\ell}(kx) + \frac{-(\ell+2)(\ell+1)}{(2\ell+1)(2\ell+3)} j_{\ell+2}.$}~\cite{Bharadwaj:2004nr}. \(W_{\nu}(x)\) is a window function spanning a particular frequency band centred on \(\nu\) accounting for finite spectral resolution, and \(x\) denotes the comoving distance along the line of sight. The window function depends on the details of the experiment; here we have assumed a Gaussian window function of width \(\Delta \nu\).
\par
The angular power spectrum of the 21 cm fluctuations at a certain frequency \(\nu\) in terms of the matter power spectrum $P_m(k)$ can then be written as,
\begin{equation}
\label{eq:Cl}
\mathcal{C}_{\ell} = \frac{2}{\pi} \int_0^{\infty} k^{2}dk P_{m}(k) \mathcal{T}^2_{\ell}(k,\nu).
\end{equation}
In order to speed up computation time, we switch to the Limber approximation~\cite{1953ApJ...117..134L,LoVerde:2008re} 
at \(\ell \geq 10^{3}\)

\section{Signatures of dark matter on 21 cm line-intensity mapping signal}
\label{sec:DMdecayannh}
Throughout this work we use the Boltzmann code \texttt{CLASS}\footnote{http://www.class-code.net}~\cite{Blas:2011rf} interfaced with the recombination code \texttt{CosmoRec}\footnote{http://www.chluba.de/CosmoRec}~\cite{Chluba:2010ca} to solve the evolution equations for \(x_e\) and \(T_{k}\) accounting for particle decay or annihilation. In this section we briefly review the standard recombination model plus the additional effects of energy injection from decay or annihilation of DM particles. We then show how the standard 21 cm global signal and angular power spectrum, described in the previous section, is affected by exotic DM energy injection through this dependence on the thermal and ionization history. 

\subsection{Exotic energy injection: thermal and ionization history and the global 21 cm signal}
\label{sec:global_T21}
Here we follow standard Peebles' recombination~\cite{Peebles:1968ja} (further developed in~\cite{Seager:1999bc, Chluba:2009uv, Chluba:2010ca, AliHaimoud:2010dx}). Energy deposited in the medium from injection of energetic particles, denoted by \( \left. \frac{dE}{dVdt}\right |_{\mathrm{dep}}\), ionizes, excites and heats the medium. The amount of energy going to each of these channels is denoted by the subscripts \(c=\{i, \alpha, h\} \), respectively. However, not all the injected energy from the decay and annihilation products is deposited in the medium. The amount of energy deposited depends strongly upon the decay and annihilation channels of the DM particle and its mass (which together determine the energy spectrum of injected particles), and also the redshift of energy injection. Following the approach of~\cite{Finkbeiner:2011dx, Slatyer:2012yq} the energy deposition rate is commonly parameterised as
\begin{equation}
\left. \frac{dE}{dVdt} \right|_{\mathrm{dep}, c}^{\mathrm{dec/ann}} = f_{c}(z) \left. \frac{dE}{dVdt} \right|_{\mathrm{inj}}^{\mathrm{dec/ann}},
\end{equation}
where \(f_{c}(z)\) is a dimensionless efficiency function parameterising the amount of energy deposited in the medium in the three different channels \(c=\{i, \alpha, h\} \). The function \(f_{c}(z)\) encapsulates the dependence on the energy spectrum of the injected particles and the redshift of injection. A common approximation assumes \textit{on-the-spot} energy deposition, i.e. the energy released from the exotic process is promptly absorbed at the same redshift. A further simplification of this is to approximate the \(f_{c}(z)\) curves by a constant efficiency factor \(f_{\mathrm{eff}}\) over the entire redshift range, i.e. the fraction of injected energy that is promptly deposited in the IGM. This simplification has been shown to be a sufficient approximation in relation to CMB analyses for DM annihilation and long-lived DM decay~\cite{Galli:2009zc, Slatyer:2012yq, Poulin:2015pna, Slatyer:2016qyl}. We choose to adopt this approximation in our analysis, foregoing a full calculation of the $f_{c}(z)$ functions, for this initial exploration of the effect of DM energy injection in the 21 cm LIM angular power spectrum. We later discuss the implications of this choice in Appendix~\ref{sec:decay_case}. 
\par
The evolution of the ionization fraction (or free electron fraction) \(x_{e}\) is governed by~\cite{Chen:2003gz},
\begin{equation}\label{eq:xe}
\frac{dx_{e}}{dz} = \frac{1}{(1+z)H(z)}[R_{s}(z) - I_{s}(z) - I_{X}(z)],
\end{equation}
with standard recombination and ionization rates respectively given by,  
\begin{equation}
R_{s}(z) = C \left [ \alpha_{H} x^{2}_{e} n_{H} \right ], \quad I_{s}(z) = C \left [ \beta_{H}(1-x_{e}) e^{-\frac{h\nu_{\alpha}}{k_{b}T_{k}}} \right ].
\end{equation}
The coefficient \(C\) takes into account the probability that an electron transitions from the \(n=2\) to the \(n=1\) state before being ionized, given by,
\begin{equation}
C = \frac{1 + K_{H}\Lambda_{H} n_{H} (1-x_{e})}{1 + K_{H}(\Lambda_{H} + \beta_{H})n_{H}(1-x_{e})},
\end{equation}
where \(K_{H}=\lambda_{\alpha}^{3}/8\pi H(z)\) describes the cosmological redshifting and escape of the Ly-\(\alpha\) photons, and \(\Lambda_{H}\)  is the vacuum decay rate of the metastable \(2s\) level (for a more detailed explanation, refer to Appendix A of~\cite{Poulin:2015pna}). 
\par
The final term of Eq.~\ref{eq:xe} \(I_{X}= I_{X_{i}} + I_{X_{\alpha}}\) is an effective ionization rate due to extra injection of energetic particles,   
split in terms of the direct ionization rate 
 and the excitation plus ionization rate 
 following~\cite{Poulin:2016anj} (see also~\cite{Chen:2003gz,Slatyer:2016qyl}).  
Therefore, the contributions to the ionization rate \(I_{X}\) in terms of the energy deposition rate from the DM decay/annihilation, are
\begin{equation}
\begin{split}
I_{X_{i}} = \left. -\frac{1}{n_{H}(z)E_{i}} \frac{dE}{dVdt} \right|_{\mathrm{dep},i}, \\
I_{X_{\alpha}} = \left. -\frac{1-C}{n_{H}(z)E_{\alpha}} \frac{dE}{dVdt} \right|_{\mathrm{dep},\alpha}, 
\end{split}
\end{equation}
where \(n_{H}(z)\) is the comoving number density of hydrogen nuclei,
and \(E_{i}\) and \(E_{\alpha}\) are the average ionization energy per hydrogen atom and the Lyman-\(\alpha\) energy respectively. 
Direct ionizations have no inhibition factor $C$, while extra excitations can only be effective with a probability $1-C$.
We neglect the effect of the extra energy injection on Helium ionization, which has been shown to be sub-dominant~\cite{Galli:2013dna,Slatyer:2015kla} and thus should not significantly affect our results. 
\par
In turn, the evolution of the kinetic gas temperature \(T_{k}\) follows 
\begin{equation}\label{eq:Tk}
\frac{dT_{k}}{dz} = \frac{1}{1+z} \left[ 2T_{k} + \gamma(T_{k}-T_{CMB}) \right] + K_{h},
\end{equation}
where \(\gamma\) is a dimensionless parameter defined as~\cite{Poulin:2016nat}
\begin{equation}
\gamma = \frac{8\sigma_{T}a_{R}T^{4}_{CMB}}{3Hm_{e}c} \frac{x_{e}}{1 + f_{He} + x_{e}} \, .
\end{equation}
Here \(\sigma_{T}\) is the Thompson cross section, \(a_{R}\) is the radiation constant, \(m_e\) the electron mass, \(c\) the speed of light and \(f_{He}\) is number of helium nuclei relative to hydrogen nuclei~\cite{Chluba:2009uv, Poulin:2015pna}. The additional heating rate term \(K_{h}\) due to the exotic energy injection is given by~\cite{Poulin:2015pna, Poulin:2016nat},
\begin{equation}
K_{h} = - \frac{2}{H(z)(1+z)3k_Bn_H(z)(1+f_{He}+ x_e)}\left. \frac{dE}{dVdt}\right|_{\mathrm{dep},h}\,,
\end{equation}
where $k_B$ is the Boltzmann constant. 

\begin{figure}[htb]
\centering
\includegraphics[width=0.9\textwidth, height=0.8\textheight]{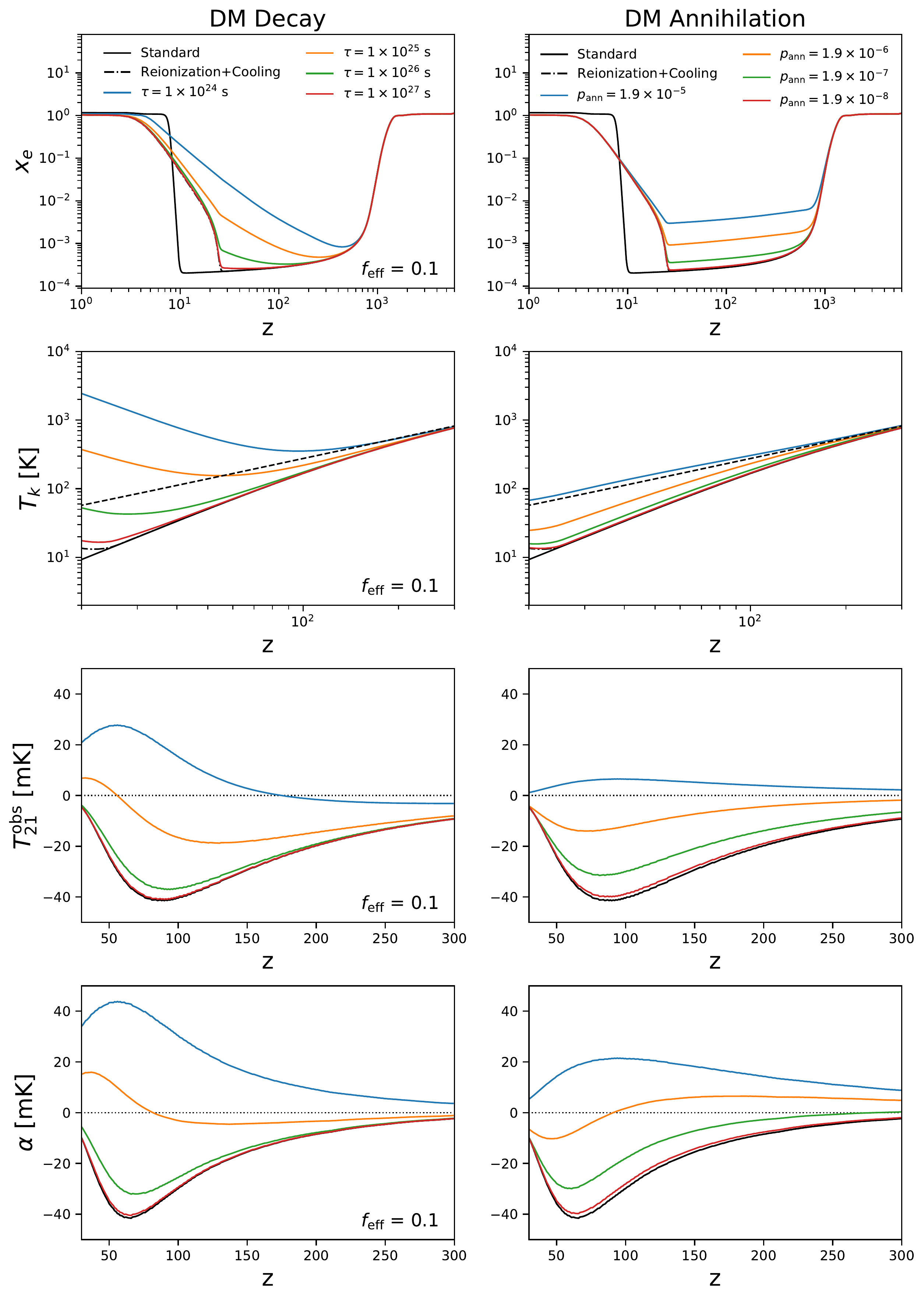}
\centering{\caption{\label{fig:Xe_Tb_T21_alp} From top to bottom: Evolution of the free electron fraction \(x_{e}\), gas kinetic temperature \(T_k\), global 21 cm differential brightness temperature \(T_{21}^{\mathrm{obs}}\) and \(\alpha(z)\) with redshift for DM decay (\textit{left}) and DM annihilation (\textit{right}). The standard \(\Lambda\)CDM prediction is shown for both a single-step reionization (black line) and with a simple stellar reionization model (black dashed) as discussed in Appendix~\ref{sec:reio}. Black dashed line in the second row represents the CMB temperature. Annihilation efficiency \(p_{\rm{ann}}\) in units of \( \mathrm{m^{3}s^{-1}kg^{-1}} \). }}
\end{figure}
While the formalism described above is general for any exotic source of particles injected to the IGM, we now focus solely on decaying and annihilating DM. The energy injection rates for DM decaying or annihilating into Standard Model particles are given by, respectively: 
\begin{equation}
\begin{split}
\left. \frac{dE}{dVdt} \right|_{\mathrm{inj}}^{\mathrm{dec}} = (1+z)^{3} f_{\mathrm{\chi}} \Omega_{\mathrm{DM}}c^{2} \rho_{c}  \frac{e^{-\frac{t}{\tau}}}{\tau}  , \\
\left. \frac{dE}{dVdt} \right|_{\mathrm{inj}}^{\mathrm{ann}} = (1+z)^{6} f_{\mathrm{\chi}}^{2}\Omega_{\mathrm{DM}}^2 c^{2} \rho_{c}^2   \frac{\langle \sigma v \rangle}{m_{\mathrm{\chi}}}
\end{split}
\end{equation}
where \(f_{\mathrm{\chi}}\) is the fraction of DM that decays/annihilates, \(\Omega_{\mathrm{DM}}\) is the present-day total abundance of cold DM, \(\rho_{c}\) is the critical density of the Universe today, \(\tau\) is the lifetime of the decaying particle (related with the decay rate $\Gamma$ by \(\tau \equiv \Gamma^{-1}\)), \(m_\mathrm{\chi}\) is the DM particle mass, and \(\langle \sigma v \rangle\) is the thermally averaged self-annihilation cross section. As customary, in these equations a smooth cosmological DM distribution is assumed, neglecting structure formation. In this work we consider only cases in which all of the DM decays/annihilates, i.e. \(f_{\mathrm{\chi}}=1\). 
\par
For the purposes of this work, for DM decay we choose to set \(f_{\rm{eff}}=0.1\) in order to explore the implications of this fiducial model on the 21 cm angular power spectrum and the potential to constrain the DM lifetime. We discuss the limitations of this choice and how our results may change quantitatively when considering more detailed injection histories in Appendix~\ref{sec:decay_case}. For DM annihilation, we can condense the DM parameters in to an effective annihilation efficiency $p_{\rm{ann}} \equiv f_{\mathrm{eff}} \langle \sigma v \rangle/m_{\mathrm{\chi}}$ and so the efficiency factor, if taken to be constant, is already absorbed into the parameter we want to constrain and we do not assume anything for \(f_{\rm{eff}}\). After the first stars form, there are other sources of heating and annihilation. We take this into account by including our simple stellar reionization model (Appendix \ref{sec:reio}). Unless otherwise stated, results presented in this paper include this stellar reionization prescription.
\par
The effect of DM decay and annihilation on the thermal and ionization history of the IGM, and the resulting 21 cm differential brightness temperature \(T_{21}\), defined in Eq.~\ref{eq:T21}, and corresponding \(\alpha(z) = \rm{d}  T_{21}/ \rm{d} \delta_b \) are shown in Figure~\ref{fig:Xe_Tb_T21_alp}. We show here the standard \(\Lambda\)CDM prediction (i.e., without exotic energy injection) both with the stellar reionization model and with the standard hyperbolic tangent transition describing reionization. For decaying DM with the considered lifetimes, the energy injection heats and ionizes the gas, resulting in an increasing ionization fraction with redshift, slowly reionizing the IGM at high redshift. Additionally, the extra energy injection heats the gas temperature \(T_{k}\), which then rises above the CMB temperature earlier than in the case of no DM decay. Furthermore, as the DM decay increases the ionization fraction, Compton scattering becomes more efficient and results in extra heating of the gas (second term of Eq. \ref{eq:Tk}) on top of the direct heating term \(K_{h}\). All of this has a strong effect in the global \(T_{21}\) signal which directly depends on the thermal history, unlike CMB. While \(T_{k} < T_{\mathrm{CMB}}\) (and collisional coupling maintains \(T_{s} < T_{\mathrm{CMB}}\)), we observe the 21 cm signal in absorption. However, due to the heating and ionization produced by the DM decays, the depth of the absorption signal reduces, even transitioning to a signal in emission for the shortest decay lifetimes considered here.
\par
The effect of annihilating DM  on the 21 cm signal is comparable, where greater annihilation efficiency leads to more ionization and heating of the gas, resulting in a shallower absorption or even emission signal in 21 cm. For annihilating DM however, the energy injection is proportional to the square of the density, hence most of the energy is injected at earlier times (before or around recombination) when the DM density was higher. The ionization fraction then freezes out at an elevated value relative to the \(\Lambda \mathrm{CDM}\) prediction, then decreases very slowly until reionization kicks in. On top of the direct heating of the gas from the energy injected in the annihilation products, the higher \(x_e\) can have the effect of delaying hydrogen-CMB decoupling (keeping \(T_k \sim T_{\mathrm{CDM}}\) for longer) resulting in an increased gas temperature at later times since there has been less time to cool adiabatically. The effect of DM annihilation on gas heating at high-\(z\) is much smaller because the majority of energy is injected while Compton scattering is still efficient enough to ensure \(T_{k} \sim T_{\mathrm{CMB}} \). Therefore, DM annihilation tends to reduce the variation of $T_{21}$ with redshift.
\par
This work being a proof-of-concept, we do not include second order effects of the extra energy injection from exotic DM in our modelling. Among these, the most important might be the potential advance of reionization, with the first stars forming earlier, due to the impact of DM decays or annihilation. This would increase the redshift at which the dark ages finish, hence keeping the observation of the dark ages out of reach for ground-based experiments. However it has been shown~\cite{Liu:2016cnk} that a contribution of more than 10\% to reionization is disfavoured for almost all DM decay and annihilation scenarios which are consistent with the CMB constraints. Nevertheless, the backreaction effect of the increased ionization level due to DM energy injection can lead to greater gas heating, especially near the end of the cosmic dark ages~\cite{Liu:2019bbm} which can in turn modify the global \(T_{21}\) signal. We also do not consider the potential Wouthuysen-Field effect coupling due to the radiation injected by the exotic DM with the IGM. We leave the study of these effects to be included in a more detailed analysis in future work. Furthermore, we do not consider any DM annihilations with velocity-suppressed cross-sections (\textit{p}--wave annihilation) or models with Sommerfeld enhanced~\cite{Hisano:2004ds} annihilation cross-sections.

\subsection{Dark Matter energy injection on the 21 cm angular power spectrum}
We compute the effect of DM decay and annihilation on the 21 cm angular power spectrum during the dark ages. We only consider here the changes in the angular power spectrum due to the modifications of $P_m(k)$, $T_{21}$, and $\alpha$. In principle, the extra energy injection would not only affect the global quantities, but also the perturbations and scale dependence of the power spectrum. We leave such study for future more detailed analyses. 
\par
\begin{figure}[t]
\centering
\includegraphics[width=1\textwidth]{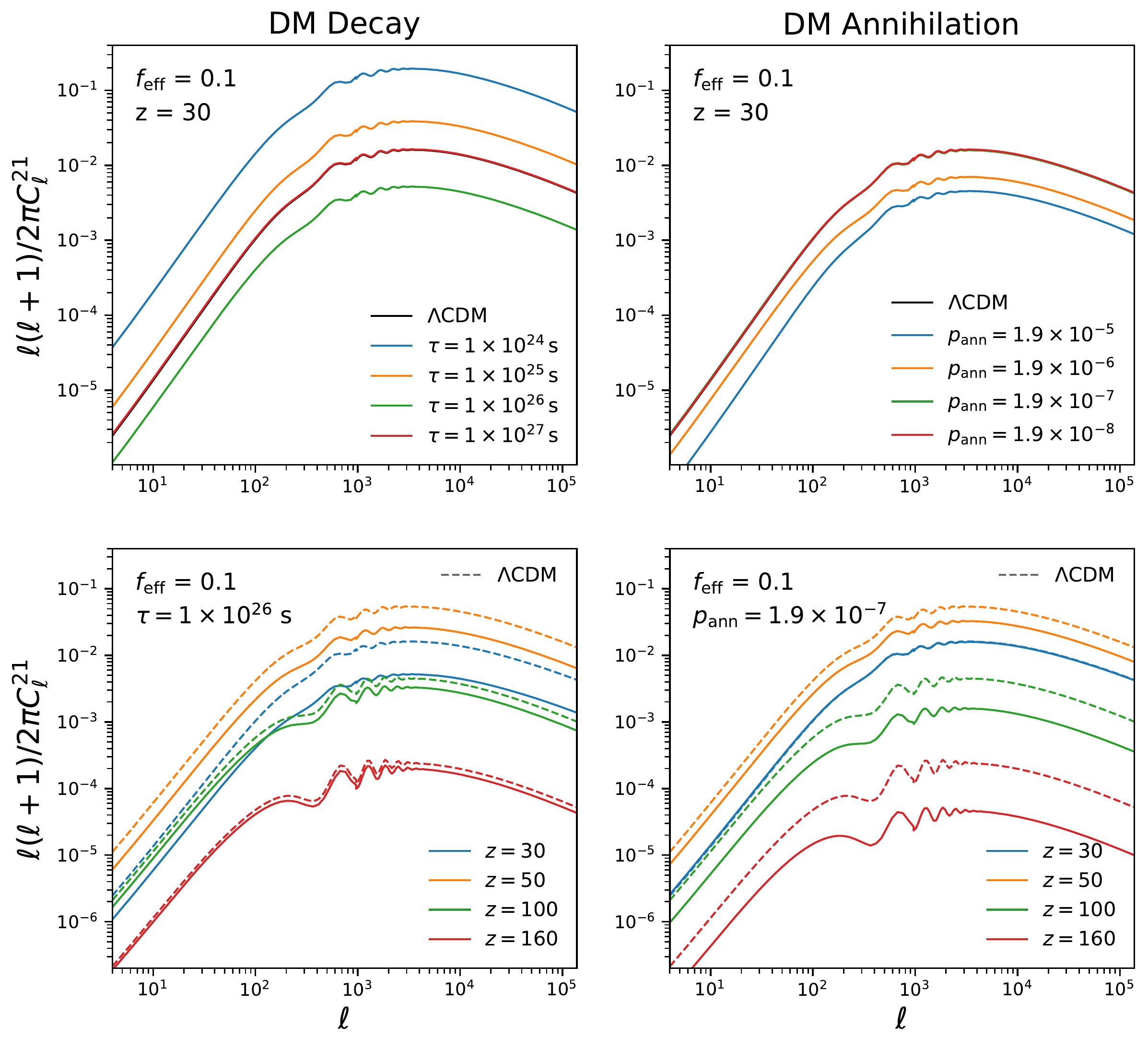}
\centering{\caption{\label{fig:21 cm_DCDM_vs_AnnDM} \textit{Top:} Angular power spectrum of fluctuations in the 21 cm brightness temperature due to energy injection from decaying (left) or annihilating (right) DM of various decay lifetimes or annihilation efficiency. \textit{Bottom:} Angular power spectrum for a fiducial DM decay of \(\tau=1 \times 10^{26} \, \mathrm{s}\) (left) and fiducial DM annihilation with \(p_{\rm{ann}} = 1.9 \times 10^{-7} \, \rm{m^3 s^{-1} kg^{-1}}\) (right) in solid lines at various redshifts during the dark ages (\(z=30-180\)), compared to the standard \(\Lambda\)CDM power spectrum at each redshift (dashed lines). Annihilation efficiency \(p_{\rm{ann}}\) in units of \( \mathrm{m^{3}s^{-1}kg^{-1}} \).}}
\end{figure}

Figure \ref{fig:21 cm_DCDM_vs_AnnDM} shows how the 21 cm angular power spectrum at \(z=30\) is modified from the standard \(\Lambda\)CDM signal by electromagnetic energy injection from either decaying or annihilating DM (top left and right panels, respectively), for several values of the decay lifetime or annihilation efficiency. The bottom left and right panels of Figure~\ref{fig:21 cm_DCDM_vs_AnnDM} show the predicted signal at various redshifts for a single fiducial decay lifetime and annihilation efficiency, respectively. The \(C_{\rm{\ell}}\)'s depend on the square of the quantities \(T_{21}\) and \(\alpha\) (Eq.~\ref{eq:dT21}--\ref{eq:Cl}), and so the power in the \(C_{\rm{\ell}}\)'s is enhanced or suppresed accordingly as the absolute value of these quantities changes.  
In other words, it is the difference in \textit{intensity} of the absorption/emission signal compared to the standard absorption signal that determines the difference on the power spectrum (regardless of whether the global \(T_{21}\) signal is in absorption or emission itself).
\par

At $z=30$,  we find an enhancement in the 21 cm fluctuation signal for decay lifetimes \(\tau \lesssim 1 \times 10^{25}\, \rm{s}\),
while for \(\tau=1 \times 10^{26}\, \rm{s}\), the extra energy injection reduces the intensity of the global absorption signal, and so
 the resulting power spectrum is suppressed compared to the \(\Lambda \rm{CDM}\) case. For even longer lifetimes, e.g., \(\tau=1 \times 10^{27}\, \rm{s}\), the global absorption signal is almost as large as the standard one, and so the power spectrum increases again to close to the \(\Lambda \rm{CDM}\) line. Within current bounds on the DM annihilation parameter (\(p_{\rm{ann}} < 1.9 \times 10^{-7} \, \rm{m^3 s^{-1} kg^{-1}} \)), the deviation from the standard signal in the angular power spectrum is much smaller than for the allowed range of decay lifetimes (\( \tau \gtrsim 1 \times 10^{25} \mathrm{s}\)) at \(z=30\). This is expected due to most of the energy injection occurring at higher redshift for annihilating DM. While the deviation for a fiducial \(p_{\rm{ann}} = 1.9 \times 10^{-7} \, \rm{m^3 s^{-1} kg^{-1}} \) is small at \(z=30\), the difference between the annihilating DM and the standard signal increases with redshift until around \(z \sim 50\) (bottom right of Figure~\ref{fig:21 cm_DCDM_vs_AnnDM}). 
 For the decaying DM lifetime \(\tau = 1 \times 10^{26} \mathrm{s}\) the signal also grows with redshift until around \(z \sim 40-50\), but then decays faster than for the annihilating DM scenario.
\par
We remind the reader that we demonstrate the effect of each decay lifetime for a fixed \(f_{\rm{eff}}=0.1\) only, and that the amount of the energy injection for a given lifetime can vary greatly with the particle mass and types of decay channels. We discuss the implications of this, exploring some examples, in Appendix~\ref{sec:decay_case}. Due to the sensitivity of the 21 cm signal to the energy injection history, a careful analysis of the effect on the 21 cm LIM signal across all decay masses and channels would allow the possibility to constrain or eventually measure DM particle properties beyond the lifetime, if potential degeneracies can be taken care of. These degeneracies are worth exploring, and we will perform such study in future research.

\subsection{DM annihilation in halos: the halo-boost}

\begin{figure}[htb]
\centering
\includegraphics[width=1\textwidth]{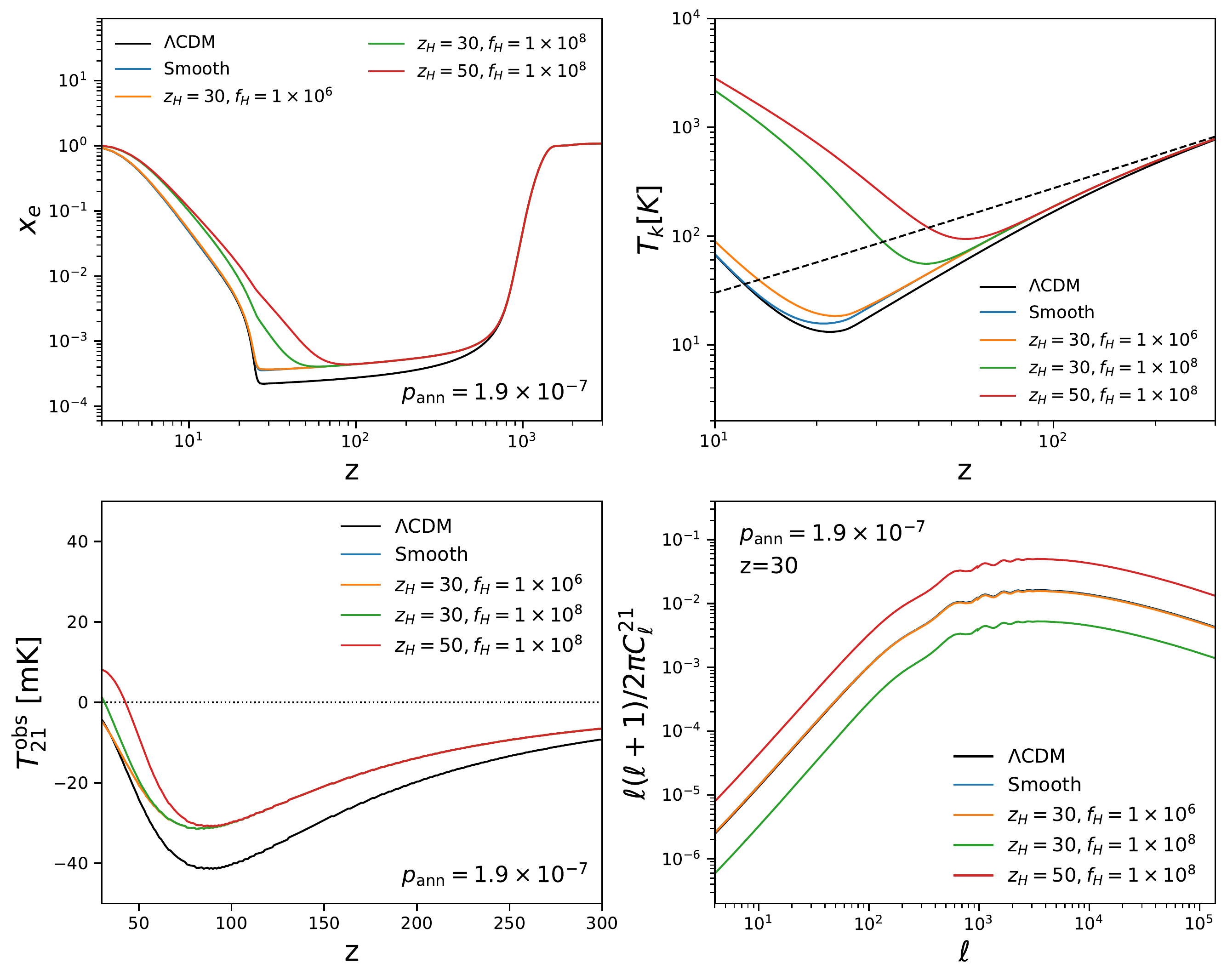}
\centering{\caption{\label{fig:21 cm_Xe_Tb_T21_Cl_halo} Evolution of free electron fraction \(x_{e}\) (\textit{top left}), gas temperature \(T_{k}\) (\textit{top right}),  and global \(T_{21}\) signal (\textit{bottom left}) with redshift, and the angular power spectrum of the 21 cm fluctuations at \(z=30\) (\textit{bottom right}), comparing the effect of energy injection of annihilating DM from the smooth background only and with the effect of various halo boost factors included. Stellar reionization model is included. Annihilation efficiency is fixed at the maximum value allowed by current Plank bounds \(p_{\rm{ann}} = 1.9 \times 10^{-7} \, \rm{m^3 s^{-1} kg^{-1}}. \)}}
\end{figure}

Below \(z \sim 100 \) the effects of structure formation may become important, enhancing the average square DM density with respect to the smooth background by some boost factor \(\mathcal{B}(z)\), such that~\cite{Huetsi:2009ex, Poulin:2015pna}:
\begin{equation}
\langle \rho^2 \rangle (z) = (1+\mathcal{B}(z)) \langle \rho \rangle^{2} (z).
\end{equation}
The energy injected from DM annihilation then becomes, 
\begin{equation}
\left. \frac{dE}{dVdt} \right|_{\mathrm{inj,smooth+halos}}^{\mathrm{ann}} = (1+z)^{6} f_{\mathrm{\chi}}^{2}\Omega_{\mathrm{DM}}^2 c^{2} \rho_{c}^2   \frac{\langle \sigma v \rangle}{m_{\mathrm{DM}}} (1+\mathcal{B}(z)).
\end{equation}
In the halo model framework, adopting the Press-Schechter formalism for the halo mass function, the boost factor is (see Appendix C of~\cite{Poulin:2015pna}), 
\begin{equation}
\mathcal{B}(z) = \frac{f_{H}}{(1+z)^{3}}\mathrm{erfc} \left ( \frac{1+z}{1+z_{H}} \right ),
\end{equation}
where \(z_{H}\) is the characteristic redshift at which halos begin to contribute, and \(f_H\) is a normalisation set by the amplitude of the boost factor at \(z=0\). According to~\cite{Poulin:2015pna} reasonable values for these parameters lie in the range \( z_H \in [20,30]\) and \( \mathcal{B}(z=0) \in [5\times 10^{4}, 10^{6}] \) (corresponding to \(f_H \in [10^{6}, 10^{8}] \)),
however this is still poorly constrained. In Figure~\ref{fig:21 cm_Xe_Tb_T21_Cl_halo} we show the effect of various halo boost parameterisations on the ionization fraction and gas temperature compared to a smooth DM distribution, and the resulting evolution of the global 21 cm temperature and its fluctuations (at \(z=30\)) for a fixed fiducial \(p_{\rm{ann}} = 1.9 \times 10^{-7} \, \rm{m^3 s^{-1} kg^{-1}} \).  Although outside the redshift range of interest here, we find that the stellar reionization model kicks in already before \(z=20\), suppressing the effect of the halo boost choice. For this reason we set \(z_{H} = 30\) and choose different fiducial values of \(f_{H}\) in our analysis. 
\par
Once the boost kicks in, it increases DM annihilation efficiency
, thereby injecting more energy, which in turn induces earlier ionization and heating of the IGM. For large enough halo boost, this can cause the gas temperature \(T_{k}\) to rise above the CMB temperature \(T_\gamma \) at early redshifts (\(z>30\)) and trigger an emission signal in the global 21 cm line during the dark ages. For smaller boost factors, the extra energy injection can result in a smaller (in absolute value) absorption/emission signal at \(z=30\) than for smooth DM, resulting in a suppression of the power spectrum with respect to the smooth DM case.


\section{Forecasts}
\label{sec:methods}
We now forecast the detectability of a decaying or annihilating DM contribution to the 21 cm angular power spectrum from the dark ages. We estimate the constraining power of a given experiment 
 using a Fisher information matrix analysis~\cite{10.2307/2342435, Tegmark:1996bz}. The Fisher information matrix is defined as
\begin{equation}
\mathcal{F}_{\alpha \beta}  = \left \langle - \frac{\partial^2 \ln \mathcal{L}(\theta)}{\partial \theta_{\alpha} \partial \theta_{\beta} } \right \rangle,
\end{equation}
where \(\mathcal{L}\) is the likelihood and \(\theta_{\alpha}\) are the cosmological model parameters.
We consider the 21 cm angular power spectrum \(C_\ell\), measured with a covariance \(\sigma^2_{C_{\ell}} \) (assumed to be diagonal), which will be given by the specific instrumental set-up and presented in Section~\ref{sec:setup} and Eq.~\ref{eq:noise}. Assuming the \(C_{\ell}\)'s are Gaussian distributed, the Fisher matrix elements corresponding to parameters \(\theta_\alpha\) and \(\theta_\beta\) can be computed from
\begin{equation}
\mathcal{F}^{21\rm{cm}}_{\alpha \beta} = \sum_{\ell,z} \frac{\partial C_{\ell}}{\partial \alpha} \frac{\partial C_{\ell}}{\partial \beta}  \sigma^{-2}_{C_{\ell\ell^\prime}}.
\end{equation}
We assume a standard \(\Lambda\)CDM plus an extra parameter to describe the DM decay or annihilation (lifetime \(\tau\) and annihilation efficiency \(p_{\rm{ann}}\) respectively) as our fiducial model, considering the following set of free parameters
\begin{equation}\label{eq:params}
\textbf{P} \equiv \{\theta_{\rm{s}}, \omega_{\rm{cdm}}, \omega_{\rm{b}}, \ln(10^{10}A_{\rm{s}} ), n_{\rm{s}}\} +  \vartheta, 
\end{equation}
where \(\theta_s\) is the angular size of the sound horizon, \(\omega_{\rm{cdm}} \equiv \Omega_{\rm{cdm}} h^2\) is the cold dark matter abundance, \(\omega_{\rm{b}} \equiv \Omega_{\rm{b}} h^2\) is the baryon abundance, \(A_{\rm{s}}\) and \(n_{\rm{s}}\) are the amplitude and spectral index of the primordial power spectrum of scalar modes respectively, and $\vartheta$ corresponds either to $\tau$ or $p_{\rm{ann}}$ depending on the DM scenario.
We adopt the latest Planck 2018 \(\rm{TT,TE, EE+low}\)-\(\rm{E}\)~\cite{Aghanim:2018eyx} best-fit values for our fiducial \(\Lambda\)CDM cosmology: \(\theta_s = 1.04109, \, \omega_{\rm{cdm}}=0.1202, \, \omega_{\rm{b}}=0.02236, \, \ln(10^{10}A_{\rm{s}}) = 3.045,  \, n_{\rm{s}}=0.9649\). To compute the final constraints we adopt the stellar reionization model discussed in Appendix~\ref{sec:reio} , and so we do not fix the redshift of reionization \(z_{\rm{reio}}\) as this is taken care of within the modelling. In our modeling, $A_{\rm{s}}$ is completely degenerate with the parameters governing the DM decay or annihilation at each redshift bin (since both control the amplitude of the power spectrum). Therefore,  we use a prior on \(A_{\rm{s}}\) from the Planck 2018 \(\rm{TT,TE, EE+low}\)-\(\rm{E}\)~\cite{Aghanim:2018eyx} results.
\par
We consider several fiducial values for \(\tau\) and \(p_{\rm{ann}}\) within the current limits set by from CMB and Big-Bang Nucleosynthesis analyses, and forecast the errors on the fiducial DM particle parameter of interest. Specifically, we forecast marginalized constraints on the DM decay lifetime for fiducial \(\tau=10^{25}-10^{27} \, \rm{s} \) and on the annihilation efficiency for fiducial \(p_{\rm{ann}}=1.9 \times (10^{-7}-10^{-8}) \, \mathrm{m^{3} s^{-1} kg^{-1}} \), still viable ranges for many DM masses and channels. We do so in order to estimate potential detections for 21 cm from the dark ages searches. For DM decay, we assume two cases for a constant $f_{\rm eff}$, with values 0.1 and 0.4, as well as time varying functions mimicking specific DM masses and decay channels. For the annihilating DM case, we consider annihilation with and without the halo boost. Regarding the parameters of the boost, we choose two configurations to bracket the reasonable range described in~\cite{Poulin:2015pna}:  Boost 1 \(z_\mathrm{H}=30, \, f_{\rm{H}} = 1 \times 10^{6}\) and Boost 2 \(z_\mathrm{H}=30, \, f_{\rm{H}} = 1 \times 10^{8}\). We do not vary the boost parameters, solely exploring the effect on the constraints for \(p_{\rm{ann}}\) for each fiducial halo boost. 
Additionally, we fix the fraction of DM that decays/annihilates to $f_{\chi}=1$, as varying this fraction would be degenerate with varying the efficiency factor $f_{\rm{eff}}$. 
\par
Finally we consider the case of non-detection, and forecast marginalized upper limits on $\tau$ and \(p_{\rm{ann}}\). We do so by assuming a fiducial model with no exotic DM and studying variations around it.

\subsection{Experimental setup}
\label{sec:setup}
The power spectrum of the instrumental noise $C^N_{\ell}$ of a radio interferometer with uniformly distributed antennas at a given frequency \(\nu\) can be written as (see e.g.~\cite{Jaffe:2000yt, Knox:2002pe,Kesden:2002ku,Zaldarriaga:2003du}),
\begin{equation}
\ell^2 C_{\ell\ell^\prime}^{N} = \frac{(2\pi)^3 T^{2}_{\mathrm{sys}}(\nu)}{\Delta \nu \, t_{o} f^{2}_{\mathrm{cover}}} \left ( \frac{\ell}{\ell_{\mathrm{cover}}(\nu)} \right )^{2}\delta^K_{\ell\ell^\prime}, 
\end{equation}
where the maximum multipole observable \(\ell_{\mathrm{cover}}(\nu) \equiv 2 \pi D_{\mathrm{base}} / \lambda(\nu) \), \(D_{\mathrm{base}}\) is the largest baseline of the interferometer, \(f_{\mathrm{cover}}\) is the fraction of this baseline covered with antennas, \(t_{\rm obs}\) is the observation time, \(T_{\mathrm{sys}}\) is the system temperature, and $\delta^K$ is the Kronecker delta. Assuming the system temperature to be the synchrotron temperature of the observed sky, we have (extrapolating to lower frequencies the results reported in Ref.~\cite{2017MNRAS.464.4995M}):
\begin{equation}
T_{\mathrm{sys}}(\nu) = 180 \left ( \frac{\nu}{180 \mathrm{MHz}}  \right )^{-2.62} \mathrm{K}.
\end{equation}
Therefore, the uncertainty in the measured 21 cm \(C_{\ell}\) at a given multipole \(\ell\) including cosmic-variance is then given by 
\begin{equation}\label{eq:noise}
\sigma_{C_{\ell\ell^\prime}} = \sqrt{\frac{2(C_{\ell} + C_{\ell\ell^\prime}^{N})^2}{f_{\mathrm{sky}} (2\ell + 1)}}\delta^K_{\ell\ell^\prime},
\end{equation}
where \(f_{\rm{sky}}\) is the fraction of sky coverage of the survey (we assume all-sky surveys, i.e.,  \(f_{\rm{sky}}=0.75 \sim 30,000 \, \rm{deg}^2\)). In addition, we consider for all cases a  frequency band \(\Delta \nu = 1 \,\mathrm{MHz}\).
\begin{table}[h!]\centering 
\ra{1.3}
\begin{tabular}{@{}rrrrcrrrcrrr@{}} \toprule
    {Spec} & {SKA} & {aSKA} & {LRA1} & {LRA2} & {LRA3} \\ \midrule
    $D_{\rm{base}}$ (km)  & 6 & 100 & 30 & 100 & 300 \\
    $f_{\rm{cover}}$  & 0.02  & 0.2 & 0.1  & 0.5 & 0.75 \\
    $t_{\rm{obs}}$ (years)  & 5  & 10 & 5  & 5 & 5 \\
    $l_{\rm{cover}}\frac{1+z}{31}$  & 5790  & 96515 & 28954 & 96515 & 289547 \\
     \bottomrule
\end{tabular}
\caption{\label{table:specs} Instrument specifications for the forthcoming SKA, an advanced ground-based SKA-like experiment (aSKA), and three realisations of a futuristic lunar radio array (LRA).}
\end{table}
We consider both the forthcoming SKA experiment, and different realisations of more advanced, futuristic earth- and lunar-based radio interferometers. 
The instrument specifications for each survey we consider are summarised in Table \ref{table:specs}. Ground-based experiments (i.e., SKA and aSKA) will be limited to observe at $z=30$, while the three realisations of the LRA will be able to perform a tomographic analysis over the redshift range \(30 \lesssim z \lesssim 200\). By considering both, we examine how much improvement in detection capabilities for a DM decay or annihilation signal is possible for different types of next-generation experiments measuring the 21 cm LIM power spectrum. Following Ref.~\cite{Munoz:2015eqa} we consider up to forty independent redshift bins.

\section{Results}
\label{sec:results}
We present forecasted constraints and lower (upper) limits on \(\tau\)  (\(p_{\rm{ann}}\)) for DM decay (annihilation), from the 21 cm LIM power spectrum as observed by the experimental setups described in Section~\ref{sec:setup}, for different versions of the SKA and Lunar radio arrays. To obtain comprehensive results, we consider two representative values of the effective efficiency parameter \(f_{\rm{eff}}=0.1, \, 0.4\) for decays, and both a smooth background and halo-boosted annihilations for the annihilating DM  scenario. We summarize our results in this section, and report all cases in detail in Tables \ref{tab:DCDM_All}, \ref{tab:AnnDM_All} and \ref{tab:upper_lower_limits} in Appendix \ref{sec:forecasts}. 
\par
We find that the forecasted errors are largely insensitive to the adopted stellar reionization model (Appendix \ref{sec:reio}) as expected, given that the reionization effect kicks in below \(z=30\). Hence, all results are reported with the stellar reionization prescription included, and we do not expect reasonable changes to the reionization model to affect our results. 
\begin{figure}[h]
\centering
\includegraphics[width=1\textwidth]{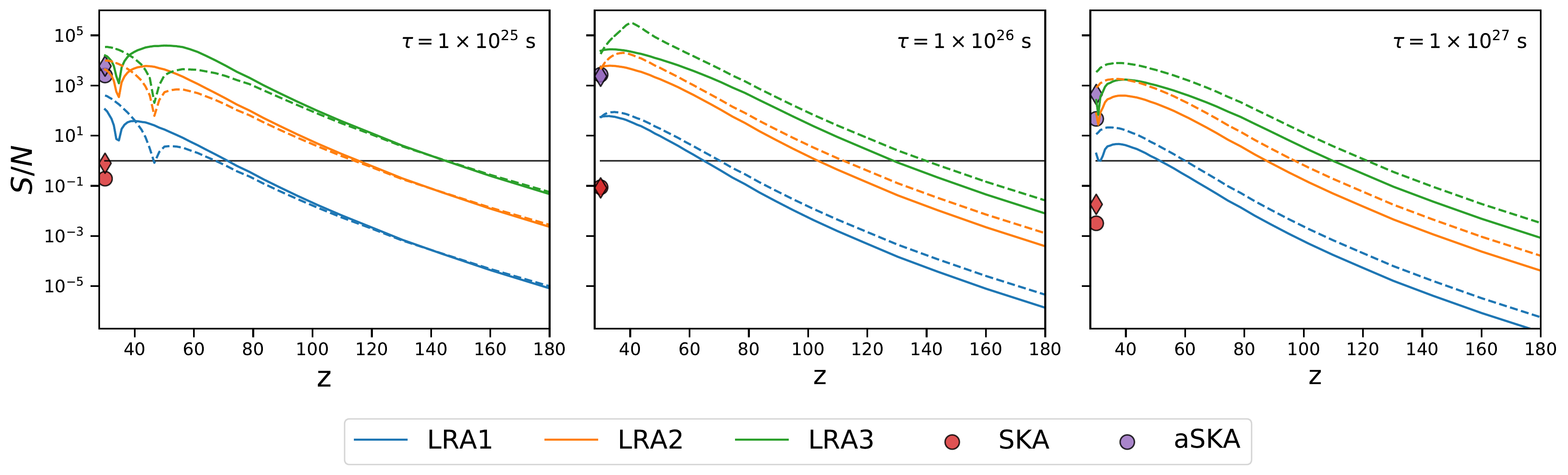}
\centering{\caption{\label{fig:21 cm_SNR} Evolution of the signal-to-noise ratio ($S/N$) with redshift between the 21 cm angular power spectrum with decaying DM lifetimes \(\tau=1\times (10^{25}-10^{27}) \, \mathrm{s}\) compared to the \(\Lambda\)CDM case (no decaying DM) as measured by SKA, an advanced SKA-like instrument (aSKA) and 3 realisations of a futuristic lunar radio array (LRA). Instrument specifications can be found in Table \ref{table:specs}. For SKA and aSKA ($z=30$), the circles represent \(f_{\mathrm{eff}}=0.1\) and diamonds represent \(f_{\mathrm{eff}}=0.4\). For LRA, solid lines represent an efficiency factor \(f_{\mathrm{eff}}=0.1\) and dashed lines represent \(f_{\mathrm{eff}}=0.4\). Solid black line represents an S/N of unity. We define $S/N = \sum_\ell(\Delta C_\ell/\sigma_{C_\ell})$. }}
\end{figure}

\begin{figure}[htb]
\centering
\includegraphics[width=1\textwidth]{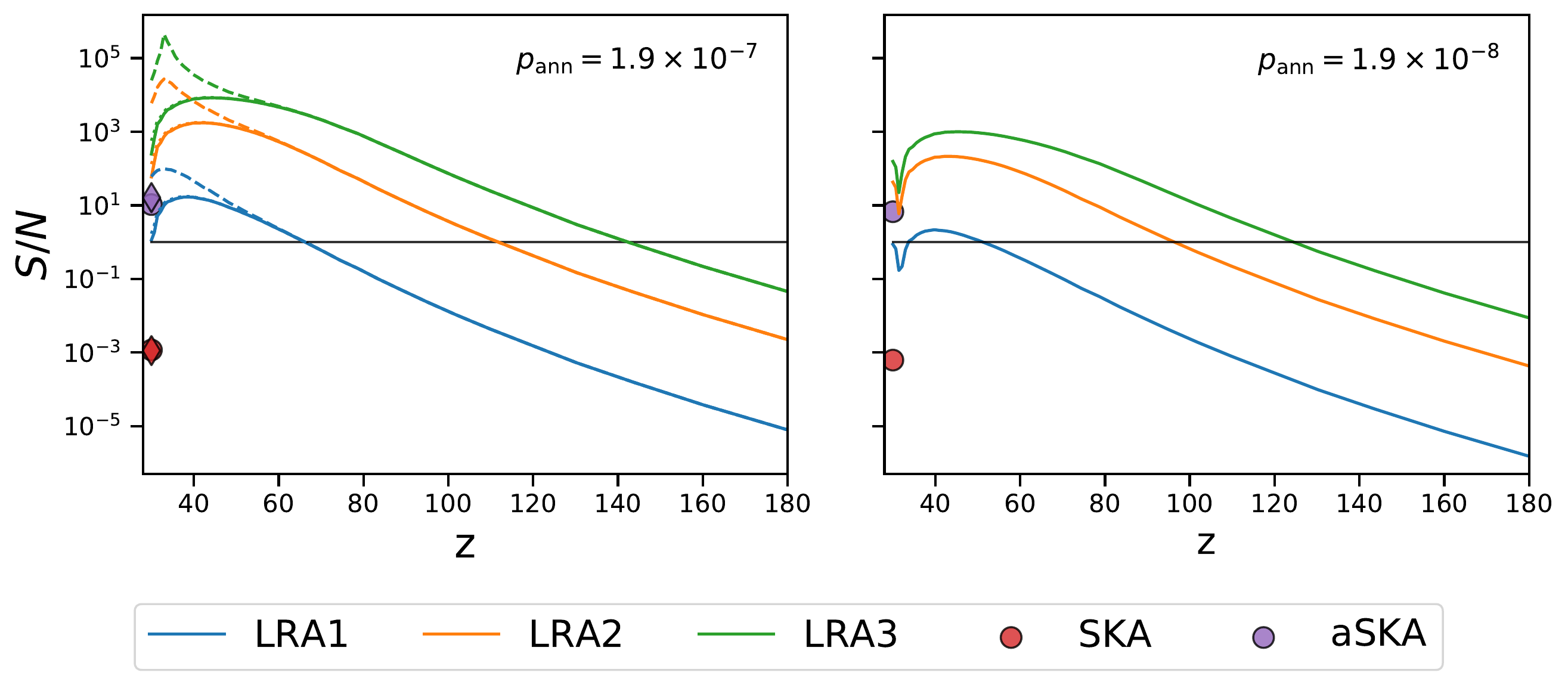}
\centering{\caption{\label{fig:21 cm_SNR_annDM} Evolution of the difference in signal-to-noise ratio ($S/N$) with redshift between 21 cm power spectrum with annihilating DM with  \(p_{\rm{ann}} = 1.9 \times (10^{-7}-10^{-8}) \mathrm{m}^{3} \mathrm{s}^{-1} \mathrm{kg}^{-1}\) in smooth background (solid), with halo boost 1 (dotted) and halo boost 2 (dashed), with respect to the \(\Lambda\)CDM case (no DM energy injection). For SKA and aSKA ($z=30$), the circles represent smooth DM distribution and diamonds represent with halo boost 1. Solid black line represents a $S/N$ of unity. $S/N$ defined as in Figure~\ref{fig:21 cm_SNR}. }}
\end{figure}

\begin{figure}[h]
\centering
\includegraphics[width=1.\textwidth]{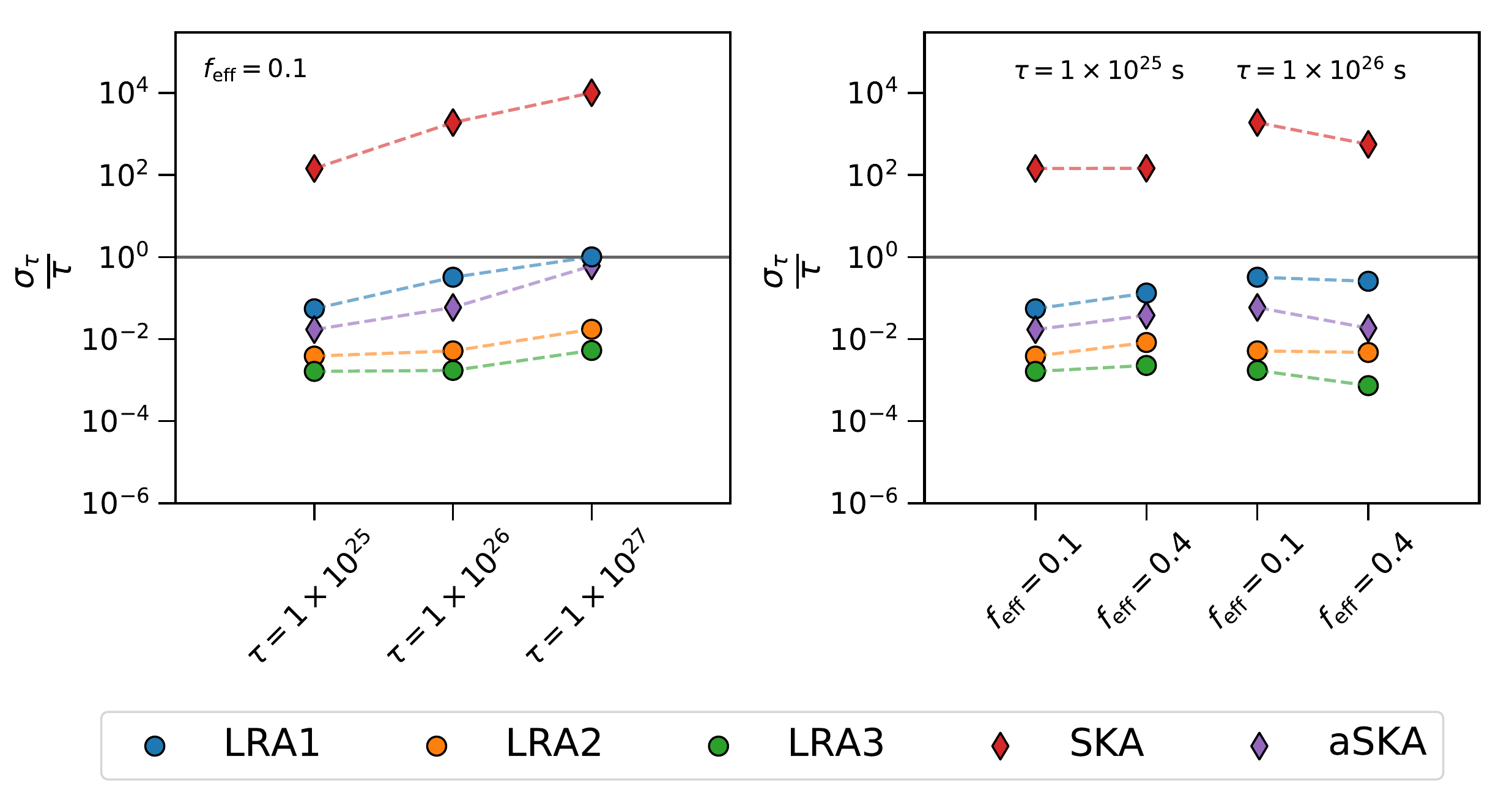}
\centering{\caption{\label{fig:Constraints_DCDM} Marginalised 68\% confidence level forecast \textit{relative errors} around each fiducial decay lifetime \(\tau=1 \times (10^{25}-10^{26}) \, \rm{s}\) (left panel) and comparing the errors around two fiducial lifetimes with both \(f_{\rm{eff}}=0.1, 0.4\) (right panel), for SKA, an advanced SKA-like instrument (aSKA) and 3 realisations of a futuristic lunar radio array (LRA). For SKA and aSKA, the constraints are based on a measurement of the 21 cm at \(z=30\) only while LRA utilises tomography of the 21 cm LIM in \(30 \leq z \leq 200\). Solid black line represents a relative error of unity.}}
\end{figure}

\begin{figure}[h]
\centering
\includegraphics[width=1.\textwidth]{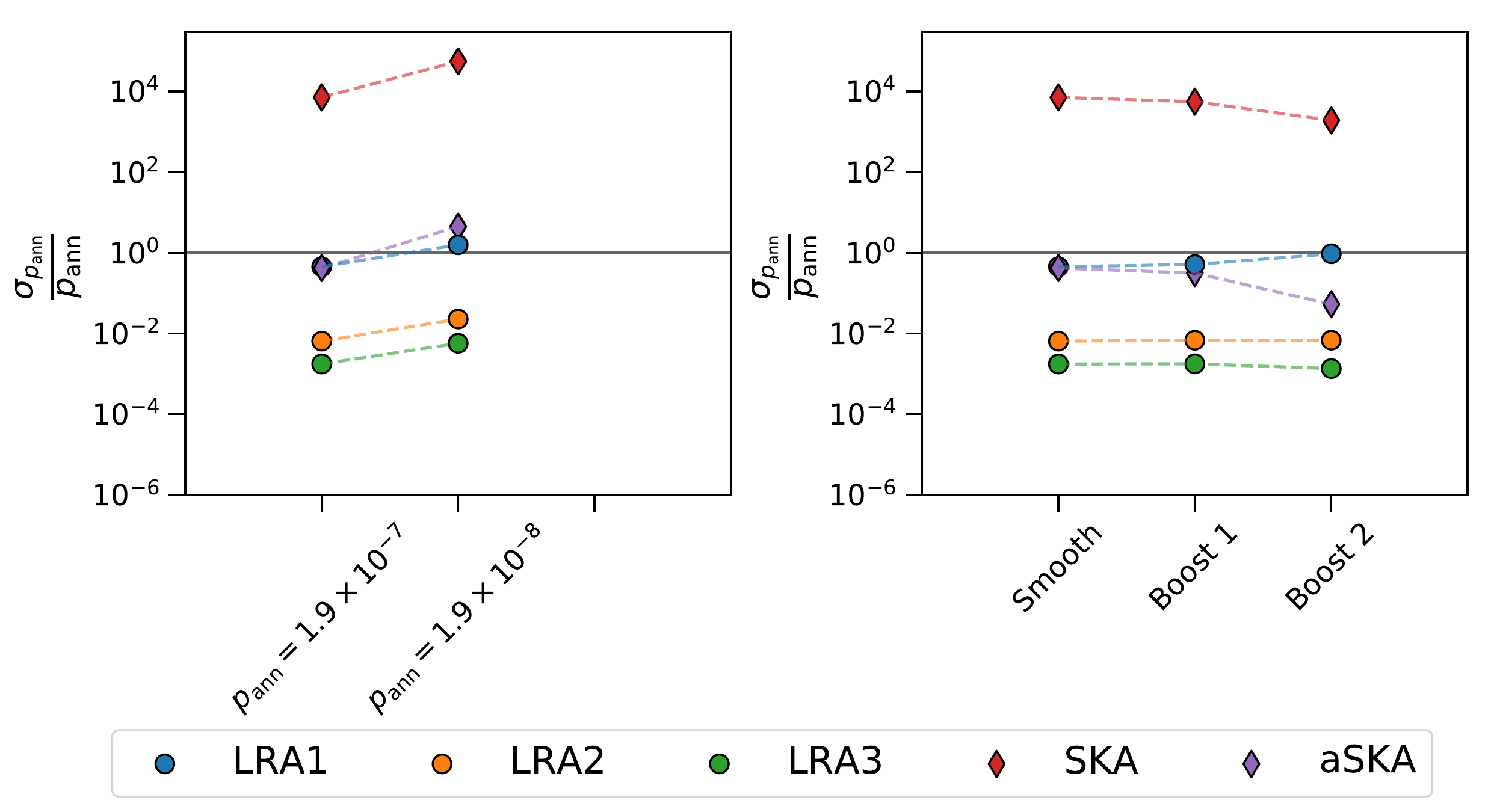}
\centering{\caption{\label{fig:Constraints_AnnDM}As in Figure~\ref{fig:Constraints_DCDM}, but here forecasted constraints are shown for the fiducial values of \(p_{\rm{ann}} = 1.9 \times (10^{-7}-10^{-8}) \rm{m}^3 s^{-1} kg^{-1}\) for a smooth DM distribution (left panel) and errors around a fixed \(p_{\rm{ann}} = 1.9 \times 10^{-7} \rm{m}^3 s^{-1} kg^{-1}\) assuming a smooth background DM distribution only compared to adding a halo boost factor to account for early structure formation (right panel). Solid black line represents a relative error of unity.}}
\end{figure}

Figures \ref{fig:21 cm_SNR} and \ref{fig:21 cm_SNR_annDM} show the evolution of the signal-to-noise ratio (\(S/N\)) with redshift of the 21 cm angular power spectrum for the fiducial decaying and annihilating DM models, respectively, as measured by SKA, aSKA (\(z=30\)) and LRA. As most of the energy injection occurs at earlier times for DM annihilation when the background DM density is higher, the \(S/N\) tends to peak at higher redshift than for decaying DM. The dip in the $S/N$ that is observed e.g. for $\tau=1 \times 10^{25}$ s is due to the crossing from absorption into emission of the global $T_{21}$ signal, resulting in a drop in signal in the $C_{\ell}$'s. Figures \ref{fig:Constraints_DCDM} and \ref{fig:Constraints_AnnDM} summarise the marginalised forecasted \textit{relative errors} around each fiducial decay lifetime and annihilation efficiency respectively, for all 5 experimental set-ups measuring the 21 cm power spectrum. 
\par
Our results demonstrate that SKA alone is unlikely to precisely constrain DM decay even for the shortest viable decay lifetimes. 
There is no case within the allowed lifetimes for which the SKA reaches a signal-to-noise ratio \(S/N\) of unity, except for decay channels with much larger energy injection, see Appendix~\ref{sec:decay_case}. 
Even under the optimistic assumption that SKA can observe up to \(z\sim35\), we found the constraints do not improve dramatically enough to enable a precise detection. With a more sensitive earth-based instrument like aSKA, it would be possible to constrain lifetimes of up to \(\tau=1 \times 10^{26}\)s with close to percent-level precision observing at $z=30$ only. Due to its large baseline, aSKA measuring the power spectrum at \(z=30\) could better constrain the DM lifetime than LRA1 even taking advantage of the full redshift range. Therefore, having a large enough baseline with enough angular resolution to resolve the information at small scales will be critical in providing precise measurements of the DM parameters. 
\par
However, only a tomographic analysis would allow the possibility to constrain the DM lifetime down to the sub-percent level. The information gain from probing higher redshifts significantly boosts the detection capabilities. While in general the $S/N$ decreases with redshift, the maximum $S/N$ appears at $z>30$ for various parameters configurations, especially as the experiment improves (see Figure \ref{fig:21 cm_SNR}). Even with a much more sensitive instrument like LRA3 (with a much larger baseline and \(f_{\rm{cover}}\)), the measurement errors at \(z=30\) alone do not improve much over what is attainable with aSKA (an improvement of around a factor of \(\sim 1.3\) at \(z=30\) compared with a factor \(\sim 11\) using the range \(30<z<200\)). As such, a lot of information is lost if these higher redshifts cannot be reached.
\par
With LRA2, it would be possible to constrain lifetimes of up to \(\tau=1 \times 10^{26}\, \rm{s}\) at the sub-percent level, already \(\sim 1\) order of magnitude below general bounds on the DM lifetime from CMB~\cite{Slatyer:2016qyl}. LRA3 could reach sub-percent level precision even for \(\tau=1 \times 10^{27}\)s, two orders of magnitude longer lifetimes than current CMB limits (given an energy injection of \(f_{\rm{eff}}=0.1\)). For decay scenarios with larger energy injection, even longer lifetimes could be probed. 
\par
We also use the Fisher matrix formalism to forecast a lower limit on the decay lifetime in case of non-detection, assuming  \(f_{\rm{eff}}=0.1\). For aSKA, we project a lower limit at \(\tau > 1.9\times 10^{27} \, \mathrm{s}\) at 95\% C.L. With a lunar-based instrument, we can reach a forecasted lower bound on the lifetime of \(\tau > 2.21 \times 10^{29}\, \mathrm{s}\) at 95\% C.L for LRA3, improving upon the current bounds from CMB measurements by more than 4 orders of magnitude (\(\tau > 1 \times 10^{25}\, \mathrm{s}\)~\cite{Slatyer:2016qyl}). A detailed report of forecasted limits for each experiment can be found in Table \ref{tab:upper_lower_limits} in Appendix \ref{sec:forecasts}. 
\par
We emphasise that the forecasts discussed above corresponding to DM decay are for two representative cases of the energy injection history, \(f_{\rm{eff}}=0.1, 0.4\). In Figure~\ref{fig:Constraints_DCDM} we can see that the relative forecasted errors do not change significantly when assuming $f_{\rm eff}=0.4$; hinting that the constraining power does not depend strongly on the value of $f_{\rm eff}$ while it is constant.
Still, we explore some specific particle decay cases (e.g. DM particles of 100 MeV decaying into electron-positron pairs, i.e., \(\chi \rightarrow e^+e^-\) and 100 GeV decay to photons, \( \chi \rightarrow \gamma \gamma\)) for a lifetime of $\tau=1 \times 10^{26}$ s and discuss the implications for the forecasted constraints in Appendix~\ref{sec:decay_case}. We find that the constraints on $\tau$ can vary by up to an order of magnitude for the specific decay models considered. The main case studied here, \(f_{\rm{eff}}=0.1\) describes reasonably well a 100 GeV decay \(\chi \rightarrow e^+e^-\). In any case, observations of the 21 cm power spectrum at \(z=30\) alone are unlikely to yield a \(S/N\) of unity for DM decay, except in decay cases where the energy injection is near maximal, e.g. MeV-scale particle decaying to electron--position pairs, or with an instrument with extremely good resolution like aSKA.
Nonetheless, 21 cm LIM tomography with LRA will have the power to improve upon CMB constraints by orders of magnitude.  
\par
As in the case of decaying DM, SKA will not reach the required sensitivity to detect an annihilating DM signal in the 21 cm angular power spectrum. Even at the upper limit set by Planck CMB measurements of \(p_{\rm{ann}} = 1.9 \times 10^{-7} \mathrm{m}^{3} \mathrm{s}^{-1} \mathrm{kg}^{-1}\) and with the largest considered halo boost, the \(S/N\) for SKA remains well below unity. aSKA may have the potential to detect a DM annihilation signal if the annihilation rate (or boost factor) is sufficiently high. In order to reach (sub-)percent level precision measurements or probe weaker annihilation rates it will be necessary to use tomographic observations of the 21 cm LIM. An instrument like LRA2 (LRA3) has the capability to constrain the DM annihilation parameter at the level of \(p_{\rm{ann}} = 1.9 \times 10^{-8} \mathrm{m}^{3} \mathrm{s}^{-1} \mathrm{kg}^{-1}\) with percent (sub-percent) precision, allowing for a very accurate measurement, at least one order of magnitude below the current limit set by Planck CMB measurements. 
\par
The importance of measuring the 21 cm power spectrum beyond \(z > 30\) to perform tomography is even more pronounced in the case of annihilating dark matter. Since the majority of energy injection occurs earlier, the contribution from exotic DM at \(z=30\) is generally significantly less than its maximum at around \(z \sim 50\). This results in a huge improvement in the forecasted measurement errors when comparing LRA constraints at \(z=30\) versus using tomography (up to 3 orders of magnitude improvement, e.g. for LRA3 measuring an annihilation rate \(p_{\rm{ann}} = 1.9 \times 10^{-8} \mathrm{m}^{3} \mathrm{s}^{-1} \mathrm{kg}^{-1}\)). 
\par
Including the effect of sub-structure in the DM distribution with the halo boost enhances the DM annihilation efficiency at low redshift, thereby boosting the \(S/N\) at later times (see Figure \ref{fig:21 cm_SNR_annDM}). 
For the most conservative boost factor we chose, Boost 1, the impact in the constraints is negligible. However with the larger Boost 2, the effect in the constraints at \(z=30\) can be significant (see right panel Figure \ref{fig:Constraints_AnnDM}). For example, there is a factor of \(\sim 8\) improvement for aSKA, and over an order of magnitude stronger constraints with LRA3 at \(z=30\). When performing tomography with LRA, the effect in the constraints is negligible. However, when including the effects of structure formation, the effective on-the-spot approximation may break down and the energy deposition curves can change substantially at low redshifts~\cite{Slatyer:2012yq}. Therefore, a full calculation of the redshift-dependent $f(z)$ (or $p_{\rm{ann}}(z)$) curves once precision 21 cm measurements are made possible should be taken into account when including halo boost effects.
\par
\begin{figure}[h]
\centering
\includegraphics[width=0.6\textwidth]{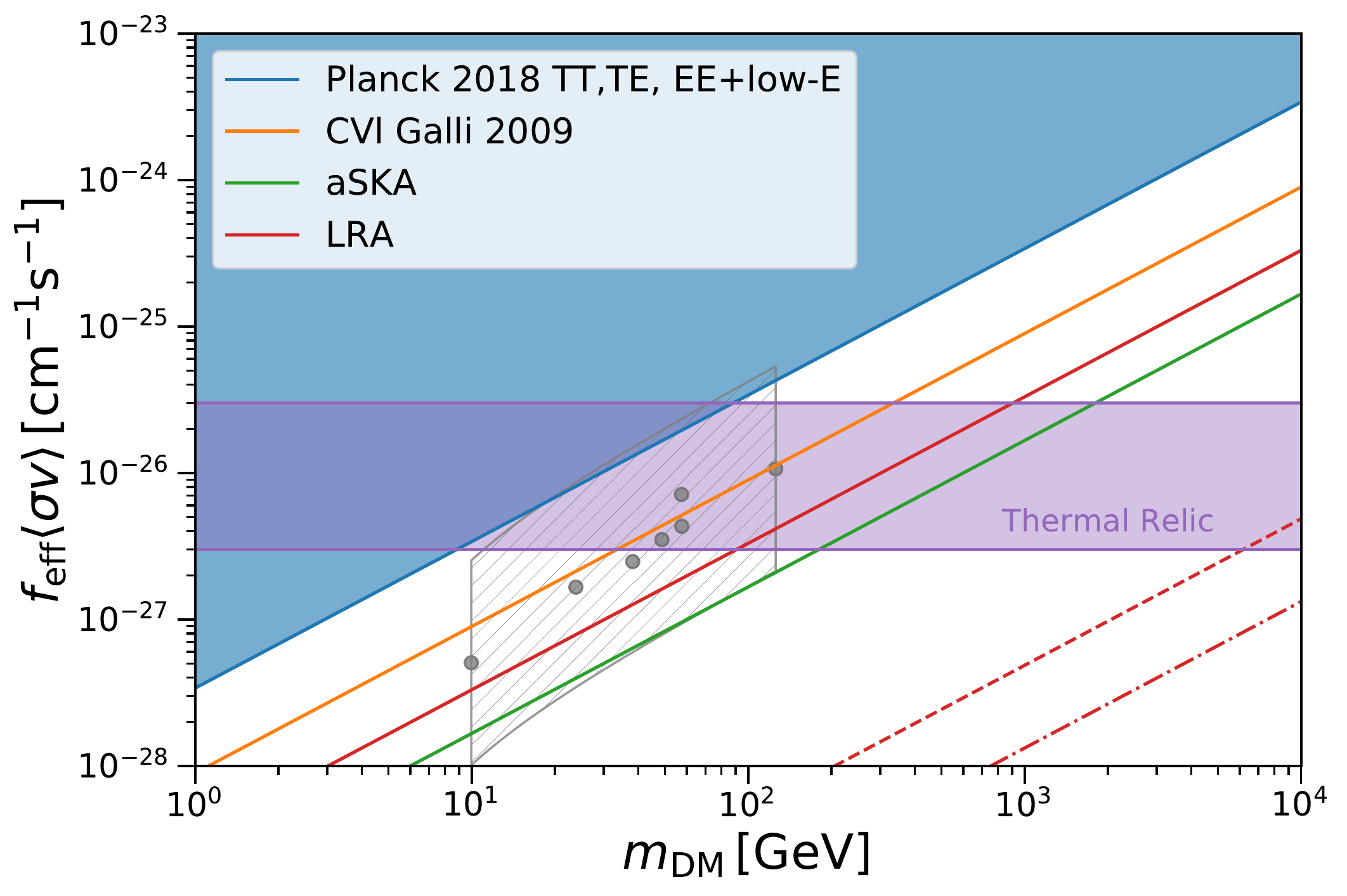}
\centering{\caption{\label{fig:mass_crossection} Projected upper limits (95\% C.L.) are shown in green for aSKA and red for LRA (solid, dashed, and dot-dashed represent LRA1--LRA3 respectively), compared to a forecasted Cosmic Variance limited CMB bound (95\% C.L.) from Ref.~\cite{Galli:2009zc} (orange). Blue shaded exclusion region is the 95\% C.L. Planck 2018 TT,TE,EE+lowE bound~\cite{Aghanim:2018eyx}. Purple horizontal band represents the thermal relic cross section of  \(\langle \sigma v \rangle = 3 \times 10^{-26} \, \mathrm{cm^3 s^{-1}}\) multiplied by a range of $f_{\rm{eff}}$ values for different annihilation channels, from \(f_{\rm{eff}}=1\) (top) to \(f_{\rm{eff}}=0.2\) (bottom). The grey circles show the best-fit DM models that are compatible with the \textit{Fermi}-LAT Galactic centre excess found in Ref.~\cite{Calore:2014nla}, and the hatched area represents the factor \(\sim 5\) uncertainty in the best-fit cross-section due to Milky Way halo parameters.}}
\end{figure}

Regarding the upper limit on the annihilation efficiency $p_{\rm ann}$ for cases of non-detection, we can compare it with the forecast from an ideal Cosmic Variance limited (CVl) CMB experiment, reported to be \(p_{\rm{ann}} < 5 \times 10^{-8} \rm{m^3 s^{-1} kg^{-1}}\) at 95\% C.L. in Ref.~\cite{Galli:2009zc}. Figure \ref{fig:mass_crossection} shows the forecasted upper limits on the annihilation efficiency for each experimental set-up, compared to the latest Planck bounds and the aforementioned CVl experiment forecast. Exploiting the power of tomography of the 21 cm LIM signal, LRA could detect or exclude values of \(p_{\rm{ann}}\) up to 1--3 orders of magnitude smaller than current CMB bounds (based on LRA1--LRA3 upper limit). Even with an earth-based instrument like aSKA, measurement of the 21 cm power spectrum can achieve 20--fold improvement over current Planck bounds with a projected upper limit of \(p_{\rm{ann}} < 9.3 \times 10^{-9} \rm{m^3 s^{-1} kg^{-1}}\) at 95\% C.L. 
\par
Assuming a thermal relic cross-section of \(\langle \sigma v \rangle \sim 3 \times 10^{-26} \, \rm{cm}^3 \rm{s}^{-1}\) and a perfect absorption efficiency of \(f_{\rm{eff}}=1\), then a non-detection of a DM annihilation signal with LRA3 would exclude a thermal WIMP of \(\mathcal{O}(10^{5})\, \mathrm{GeV} \) mass. For a more conservative choice of \(f_{\rm{eff}}=0.2\) (relevant for many annihilation channels~\cite{Madhavacheril:2013cna}), LRA3 would have the potential to probe the region of the thermal relic annihilation cross-section for \(\mathcal{O}(10^{4})\, \mathrm{GeV} \) WIMP masses. Even less powerful instruments such as aSKA and LRA1 would permit the possibility to probe a thermal relic WIMP of mass \(\mathcal{O}(10^{2}-10^{3})\, \mathrm{GeV} \), allowing to explore much of the parameter space often invoked if DM annihilations are to explain the observed \textit{Fermi}-LAT Galactic centre gamma-ray excess~\cite{Calore:2014nla} and the AMS anti-proton excess~\cite{Cholis:2019ejx}. With more sensitive instruments like LRA2 or LRA3, this region could be decisively excluded. We leave for future work a detailed computation of the energy injection histories in order to forecast constraints on the cross-section in terms of specific annihilation channels. 

\section{Discussion and Conclusions}
In this paper we propose the use of the 21 cm line-intensity mapping angular power spectrum from the dark ages to constrain decaying and annihilating dark matter through its exotic energy injection. While remaining agnostic about the specifics of the microscopic description of the DM particles (i.e., mass, branching ratios), we model generic energy injection from DM on the thermal and ionization history of the intergalactic medium, and compute how this in turn modifies the global \(T_{21}\) and the angular power spectrum of its fluctuations during the cosmic dark ages, the epoch following the last scattering of the CMB until the formation of the first stars and galaxies.
\par
We estimate the potential of next-generation and futuristic radio arrays to detect signatures of decaying or annihilating DM. Considering several fiducial cases for the DM parameters allowed by current bounds, we forecast the observational errors and limits (in the case of non-detection) on the decay lifetime \(\tau\) or annihilation efficiency \(p_{\rm{ann}}\) with the forthcoming SKA, an idealised earth-based SKA-like instrument (aSKA), and three realisations of a futuristic radio array on the lunar far-side (LRA), finding a dramatic improvement upon current and maximal constraints from other probes, such as the CMB power spectrum, for experiments beyond SKA. This is especially true for the LRA realisations, which will allow tomographic analyses of the 21 cm LIM power spectra from the dark ages. 
\par
Ground-based experiments will need exquisite angular resolution to access very small scales and be able to efficiently constrain these DM models. 
However, the real boost in constraining power comes from the tomographic analyses: in many cases the \(S/N\) remains larger than unity over a large redshift range. The information available thanks to observations at $z>30$ is most apparent for the annihilating DM case since the majority of the energy injection occurs earlier, boosting the \(S/N\) at higher redshift. In addition, measuring the signal across a wide redshift range will also be important for understanding the evolution of the signal, which may prove to be crucial for constraining the precise particle properties and breaking otherwise degenerate signals at a single redshift.
\par
Since the 21 cm signal is sensitive to both the thermal and ionization history of the IGM, it holds great promise, and has the potential to constrain much smaller or later-time energy injection from DM than the CMB, even when considering next generation (e.g.~CMB Stage-4~\cite{Abazajian:2016yjj}) or a Cosmic Variance limited CMB experiment, which is sensitive only to the ionization fraction around the time of recombination. Our results demonstrate that an instrument like LRA3 could be sensitive to annihilation efficiencies up to \( p_{\rm{ann}} = 7.4 \times 10^{-11} \rm{m}^3 s^{-1} kg^{-1}\) (95\% C.L.), reaching over 2 orders of magnitude greater sensitivity than what is ultimately accessible with an idealised CMB experiment, and over 3 orders of magnitude more than current Planck bounds, allowing to probe the thermal relic cross-section for \(\sim \mathcal{O}(10^{4})\, \mathrm{GeV}\) WIMP masses. In the case that all of the DM decays, the proposed instruments (aSKA--LRA3) could probe lifetimes up to \(\tau \sim 10^{27}-10^{29}\, \mathrm{s}\), improving by 2--4 orders of magnitude the current bounds on DM lifetime from CMB measurements. This represents an interesting region of parameter space for certain light decaying DM models (e.g. sterile nuetrinos, axinos) which are expected to have a lifetime on the order of \(\sim10^{27} \,\rm{s} \) if they are to explain the 3.5 keV photon line observed in several galaxy clusters~\cite{Bulbul:2014sua, Boyarsky:2014jta, Kong:2014gea, Adhikari:2016bei, Cosme:2017cxk}. 
\par
Given the proof-of-concept character of this work, our modelling presents several caveats due to some simplifying assumptions which should be taken into consideration. First, we assume that the extra energy injection does not accelerate the process of formation of the first stars, and that the end of the dark ages is fixed at $z\sim 30$, no matter the lifetime or annihilation rate of the DM. Neglecting the backreaction effect of the increased ionization level on the evolution of the thermal and ionization history may not be a valid assumption at the end of the cosmic dark ages~\cite{Liu:2019bbm}. Including this effect leads to greater energy deposition in heating the gas temperature \(T_k\),  which can result in 10\%-50\% stronger constraints from the global \(T_{21}\) signal~\cite{Liu:2019bbm}.  Hence by ignoring this effect we expect our forecasts to be more conservative. Second, we neglect any effect of the exotic energy injection on the spatial perturbations of the 21 cm signal, beyond its effect on the global $T_{21}$, as well as any potential coupling to the gas through the Wouthuysen-Field effect. In addition, for most of our results we assume no knowledge of the particle decay (or annihilation) process, and adopt a constant on-the-spot energy injection efficiency. The actual energy injection history is highly dependent on the particle mass and decay channel, as well as the lifetime, and so the DM lifetime forecasts presented here may change quantitatively when taking this into account. Likewise, the use of a constant efficiency factor \(f_{\rm{eff}}\) may not be a sufficient assumption during the redshift range of interest. We explore more realistic scenarios in Appendix~\ref{sec:decay_case}, and find that in these cases the significant improvement on exotic DM constraints from 21 cm LIM during the dark ages with respect to any other probe holds true. The fact that the results are sensitive to the precise energy injection history demonstrates that the 21 cm power spectrum from the dark ages holds promise of being a powerful tool to not only detect a decay or annihilation signal, but to probe the precise microscopic nature of the DM particle. A more detailed analysis, computing the corresponding \(f(z)\) functions over the whole redshift range to specific microscopic DM descriptions (using e.g.  \texttt{DarkHistory}~\cite{Liu:2019bbm} or \texttt{ExoCLASS}~\cite{Stocker:2018avm}), ascertaining whether different options could be distinguishable, is left for future work. 
Finally, the forecasted constraints presented here assume that the relevant foregrounds can be characterised and subtracted perfectly well, and we consider only the cosmic variance and instrumental noise in our final forecasted uncertainties in order to evaluate the potential reach of future 21 cm observations.
\par
Observing a deviation from the standard \(\Lambda \mathrm{CDM}\) signal in the 21 cm global sky-averaged brightness temperature or its fluctuations during the dark ages would be strongly indicative of exotic physics, as there is no known astrophysical process which could mimic such a signal. 
However, in this work we have not considered alternative sources of exotic energy injection during the dark ages other than DM decays or annihilations, which could be degenerate with the studied DM models. Nevertheless, the signatures from decaying and annihilating DM on the 21 cm angular power spectrum is qualitatively rather different from that expected due to e.g.~a population of primordial black holes~\cite{Bernal:2017nec} which enhances the power spectrum significantly at small scales only. Similarly, models with DM--baryon scattering are expected to have a distinct effect in the 21 cm signal~\cite{Munoz:2015bca, Tashiro:2014tsa}. Therefore we expect that different signatures from different exotic phenomena will be distinguishable from the DM models discussed in this work. 
\par
Another important effect to consider is that of massive neutrinos, which suppress the growth of perturbations (at small scales) in a scale-dependent and redshift dependent manner, in turn affecting the 21 cm power spectrum of fluctuations. The 21 cm window promises to be highly complementary to the CMB and LSS windows used thus far to constrain neutrino properties, due to the different redshift range and the sheer number of modes accessible. Hence, the observational efforts examined in this work may at the same time provide a powerful probe of neutrino properties.  
\par
In summary, if DM is coupled to the visible sector beyond gravitational interactions, then the 21 cm LIM signal will be very sensitive to exotic energy injection from its interactions with Standard Model particles. Measuring the signal during the cosmic dark ages can provide a clean and robust measurement of the DM properties, free from complication due to astrophysical processes. We show how the 21 cm angular power spectrum will be an extremely powerful probe of the DM decay and annihilation rate, with the potential to measure orders of magnitude weaker signals than those accessible with current or even idealised CMB experiments. At the same time, the potential constraints are complementary with indirect detection probes (or even stronger e.g. MeV--GeV scale particles decaying to \(e^+e^-\) pairs), while being more robust to uncertainties in astrophysical modelling, especially at $z>30$. While we highlight the power of 21 cm LIM in the dark ages to measure DM annihilation and decay, opening a window to this unexplored epoch in cosmic history will undoubtedly bring other major findings and will be highly complementary to forthcoming experiments aiming to measure the global 21 cm signal and the power spectrum from the epoch of reionization and cosmic dawn.

\acknowledgments

Funding for this work was partially provided by the Spanish MINECO under projects AYA2014-58747-P AEI/FEDER UE and MDM-2014-0369 of ICCUB (Unidad de Excelencia Maria de Maeztu). KS has received funding from the European Union's Horizon 2020 research and innovation
programme under the Marie Sk\l{}odowska-Curie grant agreement No. 713673. KS has received financial support through the ``la Caixa" INPhINIT Fellowship Grant for Doctoral studies at Spanish Research Centres of Excellence, ``la Caixa" Banking Foundation, Barcelona, Spain.
JLB has been supported by the Spanish MINECO under grant BES-2015-071307, co-funded by the ESF, during part of the development of this project. AR has received funding from the People Programme (Marie Curie Actions) of the European Union H2020 Programme under REA grant agreement number 706896 (COSMOFLAGS). LV acknowledges support of  European Union's Horizon 2020 research and innovation programme ERC (BePreSySe, grant agreement 725327). JC is supported by the Royal Society as a Royal Society URF at the University of Manchester, UK and 
the ERC Consolidator Grant {\it CMBSPEC} (No.~725456) as part of the European Union's Horizon 2020 research and innovation program.

\appendix

\section{Reionization modelling}
\label{sec:reio}
In this work we focus on the dark ages before the formation of the first stars, but given the large uncertainties around the reionization epoch, we consider how the modelling of the reionization might affect the 21 cm signal at \(z \gtrsim 30\) and thereby our results.  While the physics of reionization is poorly constrained, a new generation of upcoming experiments such as the Hydrogen Epoch of Reionization Array (HERA~\cite{DeBoer:2016tnn}),  James Webb Space Telescope (JWST~\cite{Gardner:2006ky}) and the Dark Ages Radio Explorer (DARE~\cite{Burns:2011wf}) will soon start to explore this period of the cosmos. To date we have only weak upper limits on the redshift of reionization and its physical processes. Nevertheless, we know the low-redshift Universe is fully ionized (by around \(z \sim 8\)), and this process is thought to be mostly due to the astrophysical processes of the first luminous sources.
\par
The standard procedure (e.g., widely assumed for CMB power spectra analyses) is to model the transition to a fully ionized Universe with  a single-step half-hyperbolic tangent function. This is unrealistic, and only modifies \(x_{e}\) while leaving the evolution of the inter galactic medium temperature unchanged without any treatment of how the astrophysical sources of reionization also heat the kinetic gas temperature \(T_{k}\). We follow the approach of~\cite{Poulin:2015pna}, adopting their simple model of stellar reionization and implementing it into \texttt{CosmoRec}. A source term is added to the \(x_{e}\) evolution equation (Eq.~\ref{eq:xe} in Section~\ref{sec:global_T21}) that accounts for Lyman continuum photons from UV sources in star-forming galaxies, thought to be a primary source of reionization. In addition to the ionizing radiation, a source term is added to the kinetic gas temperature evolution equation (Eq.~\ref{eq:Tk} in Section~\ref{sec:global_T21}) to account for the extra heating of the intergalactic medium due to e.g. X-rays from the stellar population. We refer the reader to~\cite{Poulin:2015pna} for a detailed description. Additional collisional cooling terms are added following~\cite{Chluba:2015lpa}. 
\par
Unless otherwise stated, all results shown in this paper are computed assuming this stellar reionization model. We found no significant change in our results from using the standard \texttt{CLASS} prescription for reionization, indicating that the 21 cm signal measured from the dark ages should be largely insensitive to the reionization process. Nonetheless, the extra ionization and heating of the IGM might advance the formation of the first stars, increasing the redshift at which the dark ages would end. We leave the exploration of the impacts of this relation for future work. 
However it has been shown that DM annihilation cannot be a dominant contribution to cosmic reionization if it is to be consistent with CMB results \cite{Poulin:2015pna}, requiring an overly large annihilation rate or halo boost. It is shown in Ref.~\cite{Liu:2016cnk} that for the most part DM cannot contribute by more than 10\% to reionization.

\section{21 cm forecast results}
\label{sec:forecasts}
\subsection{Decaying DM}
\begin{table}[]
\vspace{5mm}
\ra{1.3}
\centering
\begin{tabular}{rcc|ccc}
\hline
\multicolumn{1}{l}{Experiment} & Redshift range                   & Efficiency factor      & \multicolumn{3}{c}{Relative Error $ \sigma_{\tau}/ \tau $ } \\
\multicolumn{1}{l}{}           & \multicolumn{1}{l}{}             & \multicolumn{1}{l|}{}  & $\tau = 1 \times 10^{25}$     & $\tau = 1 \times 10^{26}$    & $\tau = 1 \times 10^{27}$    \\ \hline
SKA                            & \multirow{2}{*}{$z=30$}          & $f_{\mathrm{eff}}=0.1$ & $1.43 \times 10^{2}$          & $1.89 \times 10^{3}$         & $1.01 \times 10^{4}$         \\
                               &                                  & $f_{\mathrm{eff}}=0.4$ & $1.45 \times 10^{2}$          & $5.56 \times 10^{2} $        & $2.3  \times 10^{3}$        \\ \hline
\multirow{2}{*}{aSKA}       & \multirow{2}{*}{$z=30$}          & $f_{\mathrm{eff}}=0.1$ & $1.70 \times 10^{-2}$         & $5.88 \times 10^{-2}$        & $6.07 \times 10^{-1}$        \\
                               &                                  & $f_{\mathrm{eff}}=0.4$ & $3.79 \times 10^{-2}$         & $1.84 \times 10^{-2}$        & $1.11 \times 10^{-1}$        \\ \hline
\multirow{4}{*}{LRA1}          & \multirow{2}{*}{$z=30$}          & $f_{\mathrm{eff}}=0.1$ & $2.42 \times 10^{-1}$         & $2.89$                       & $1.65 \times 10^{1}$         \\
                               &                                  & $f_{\mathrm{eff}}=0.4$ & $2.75 \times 10^{-1}$         & $8.63 \times 10^{-1}$        & $3.62$                       \\
                               & \multirow{2}{*}{$30 < z < 200 $} & $f_{\mathrm{eff}}=0.1$ & $5.46 \times 10^{-2}$         & $3.21 \times 10^{-1}$        & $1.0$                       \\
                               &                                  & $f_{\mathrm{eff}}=0.4$ & $1.32 \times 10^{-1}$         & $2.56 \times 10^{-1}$        & $1.11$                       \\ \hline
\multirow{4}{*}{LRA2}          & \multirow{2}{*}{$z=30$}          & $f_{\mathrm{eff}}=0.1$ & $1.52 \times 10^{-2}$         & $3.17 \times 10^{-2}$        & $4.04 \times 10^{-1}$        \\
                               &                                  & $f_{\mathrm{eff}}=0.4$ & $3.52 \times 10^{-2}$         & $1.04 \times 10^{-2}$        & $7.19 \times 10^{-2}$        \\
                               & \multirow{2}{*}{$30 < z < 200 $} & $f_{\mathrm{eff}}=0.1$ & $3.84 \times 10^{-3}$         & $5.15 \times 10^{-3}$        & $1.73 \times 10^{-2}$        \\
                               &                                  & $f_{\mathrm{eff}}=0.4$ & $8.24 \times 10^{-3}$         & $4.71\times 10^{-3}$        & $1.18 \times 10^{-2}$        \\ \hline
\multirow{4}{*}{LRA3}          & \multirow{2}{*}{$z=30$}          & $f_{\mathrm{eff}}=0.1$ & $1.41 \times 10^{-2}$         & $1.34 \times 10^{-2}$        & $2.60 \times 10^{-1}$        \\
                               &                                  & $f_{\mathrm{eff}}=0.4$ & $3.18 \times 10^{-2}$         & $5.55 \times 10^{-3}$        & $4.70 \times 10^{-2}$        \\
                               & \multirow{2}{*}{$30 < z < 200 $} & $f_{\mathrm{eff}}=0.1$ & $1.63 \times 10^{-3}$         & $1.72 \times 10^{-3}$        & $5.28 \times 10^{-3}$        \\
                               &                                  & $f_{\mathrm{eff}}=0.4$ & $2.27 \times 10^{-3}$         & $7.32 \times 10^{-4}$        & $2.91 \times 10^{-3}$        \\ \hline
\end{tabular}
\caption{Forecasted 68\% C.L. relative errors on the lifetime $\tau$ for each fiducial parameterisation and experimental set-up measuring the 21 cm angular power spectrum.}
\label{tab:DCDM_All}
\end{table}

\newpage
\subsection{Annihilating DM}

\begin{table}[h!]
\centering
\vspace{5mm}
\ra{1.3}
\begin{tabular}{rcc|cc}
\hline
\multicolumn{1}{l}{Experiment} & Redshift range                   & Boost Factor                     & \multicolumn{2}{c}{Relative Error $ \sigma_{p_{\rm{ann}}}/ p_{\rm{ann}}$} \\
\multicolumn{1}{l}{}           & \multicolumn{1}{l}{}             &                                  & $p_{\rm{ann}} = 1.9 \times 10^{-7}$ & $p_{\rm{ann}} = 1.9 \times 10^{-8}$ \\ \hline
\multirow{3}{*}{SKA}           & \multirow{3}{*}{$z=30$}          & Smooth                           & $7.07 \times 10^{3}$                & $5.57 \times 10^{4}$                \\
                               &       -                          & $z_h = 30, \, f_h=1 \times 10^6$ & $5.54 \times 10^{3}$                &                     -                \\
                               &                                  & $z_h = 30, \, f_h=1 \times 10^8$ & $1.9 \times 10^{3}$                 &                      -               \\ \hline
\multirow{3}{*}{aSKA}       & \multirow{3}{*}{$z=30$}          & Smooth                           & $4.18 \times 10^{-1}$               & $4.47$                              \\
                               &                                  & $z_h = 30, \, f_h=1 \times 10^6$ & $3.13 \times 10^{-1}$               &            -                         \\
                               &                                  & $z_h = 30, \, f_h=1 \times 10^8$ & $5.32 \times 10^{-2}$               &            -                         \\ \hline
\multirow{6}{*}{LRA1}          & \multirow{3}{*}{$z=30$}          & Smooth                           & $11.33$                             & $9.35 \times 10^{1}$                \\
                               &                                  & $z_h = 30, \, f_h=1 \times 10^6$ & $8.82$                              &                 -                    \\
                               &                                  & $z_h = 30, \, f_h=1 \times 10^8$ & $2.93$                              &                -                     \\
                               & \multirow{3}{*}{$30 < z < 200 $} & Smooth                           & $4.50 \times 10^{-1}$               & $1.58$                              \\
                               &                                  & $z_h = 30, \, f_h=1 \times 10^6$ & $5.11 \times 10^{-1}$               &          -                           \\
                               &                                  & $z_h = 30, \, f_h=1 \times 10^8$ & $9.42 \times 10^{-1}$               &           -                          \\ \hline
\multirow{6}{*}{LRA2}          & \multirow{3}{*}{$z=30$}          & Smooth                           & $2.91 \times 10^{-1}$               & $3.39$                              \\
                               &                                  & $z_h = 30, \, f_h=1 \times 10^6$ & $2.41 \times 10^{-1}$               &             -                        \\
                               &                                  & $z_h = 30, \, f_h=1 \times 10^8$ & $2.65 \times 10^{-2}$               &             -                        \\
                               & \multirow{3}{*}{$30 < z < 200 $} & Smooth                           & $6.45 \times 10^{-3}$               & $2.28 \times 10^{-2}$               \\
                               &                                  & $z_h = 30, \, f_h=1 \times 10^6$ & $6.78 \times 10^{-3}$               &              -                       \\
                               &                                  & $z_h = 30, \, f_h=1 \times 10^8$ & $6.80 \times 10^{-3}$               &              -                       \\ \hline
\multirow{6}{*}{LRA3}          & \multirow{3}{*}{$z=30$}          & Smooth                           & $2.15 \times 10^{-1}$               & $2.37$                              \\
                               &                                  & $z_h = 30, \, f_h=1 \times 10^6$ & $1.54 \times 10^{-1}$               &               -                      \\
                               &                                  & $z_h = 30, \, f_h=1 \times 10^8$ & $7.88 \times 10^{-3}$               &                -                     \\
                               & \multirow{3}{*}{$30 < z < 200 $} & Smooth                           & $1.75 \times 10^{-3}$               & $5.70 \times 10^{-3}$               \\
                               &                                  & $z_h = 30, \, f_h=1 \times 10^6$ & $1.76 \times 10^{-3}$               &                -                     \\
                               &                                  & $z_h = 30, \, f_h=1 \times 10^8$ & $1.35 \times 10^{-3}$               &                -                     \\ \hline
\end{tabular}
\caption{Forecasted 68\% C.L. relative errors on the annihilation efficiency $p_{\rm{ann}}$ for each fiducial parameterisation and experimental set-up measuring the 21 cm angular power spectrum.}
\label{tab:AnnDM_All}
\end{table}

\FloatBarrier
\newpage
\subsection{Non-detection limits} 
The upper limits for the decay lifetime are only valid for an energy injection of \(f_{\mathrm{eff}}=0.1\) (which roughly corresponds to 100 GeV decay to electron-positrons). 
\begin{table}[h!]
\centering
\begin{tabular}{@{}rcc@{}}\toprule
Experiment & $\tau$ 95\% C.L.  & $p_{\mathrm{ann}}$ 95\% C.L.  \\ \midrule
SKA & $> 9.7 \times 10^{22} $ s & $<  1.9 \times 10^{-4} \, \mathrm{m}^{3} \mathrm{s}^{-1} \mathrm{kg}^{-1}$ \\
aSKA& $> 1.9 \times 10^{27}$ s & $< 9.3 \times 10^{-9}\, \mathrm{m}^{3} \mathrm{s}^{-1} \mathrm{kg}^{-1}$ \\
LRA1 & $> 1.3 \times 10^{27}$ s & $< 1.9 \times 10^{-8}\, \mathrm{m}^{3} \mathrm{s}^{-1} \mathrm{kg}^{-1}$ \\
LRA2 & $> 7.4 \times 10^{28}$ s & $< 2.7 \times 10^{-10}\, \mathrm{m}^{3} \mathrm{s}^{-1} \mathrm{kg}^{-1}$ \\
LRA3 & $> 2.2 \times 10^{29}$ s & $< 7.4 \times 10^{-11}\, \mathrm{m}^{3} \mathrm{s}^{-1} \mathrm{kg}^{-1}$ \\
\bottomrule
\end{tabular}
\caption{Forecasted 95\% C.L. lower (upper) bounds on the DM decay lifetime (annihilation efficiency) for each experimental set-up measuring the 21 cm angular power spectrum.}
\label{tab:upper_lower_limits}
\end{table}

\section{Comparison to specific decay channels}
\label{sec:decay_case}
We compare our generic approach of a constant effective efficiency parameter using the on-the-spot treatment to a more precise energy injection history for a few specific decay masses and channels. For the chosen decay channels and masses, we fit \(f_{\rm{eff}}(z)\) to the predictions from~\cite{Slatyer:2012yq}.
\begin{figure}[ht]%
    \centering
    \subfloat{\includegraphics[width=1.\textwidth]{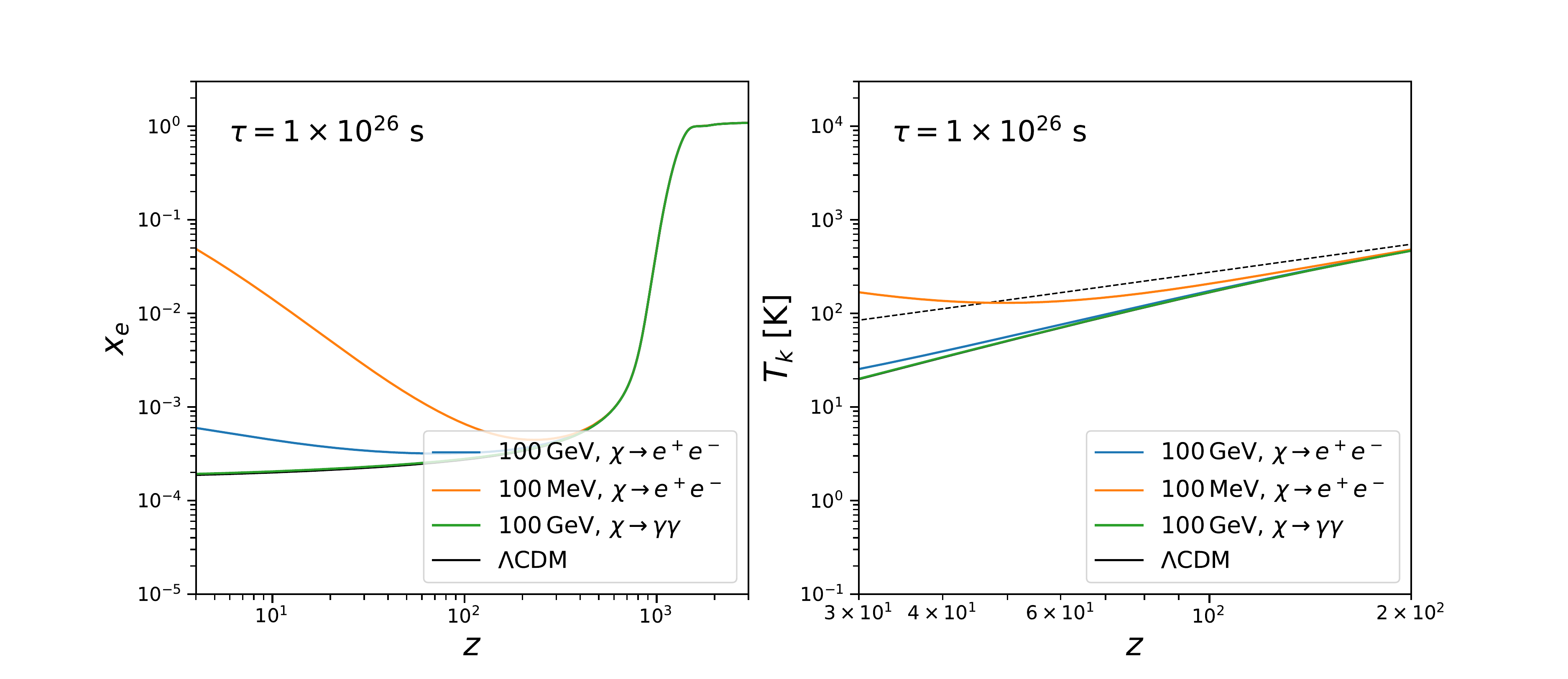} }%
    \vskip\baselineskip
    \vspace{-2.5em}
     \centering
    \subfloat{\includegraphics[width=1.\textwidth]{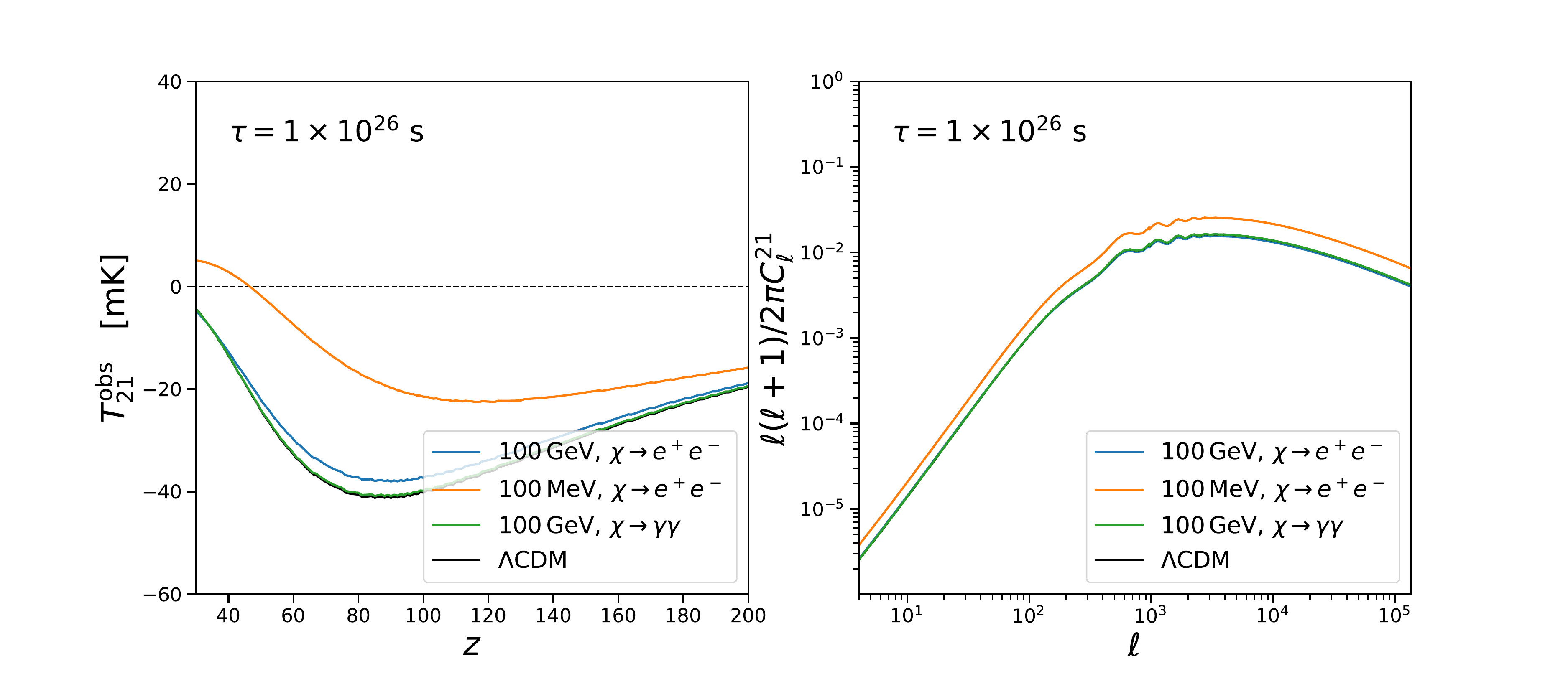} }%
\caption{Evolution of free electron fraction \(x_{e}\) (\textit{top left}), gas temperature \(T_{k}\) (\textit{top right}),  and global \(T_{21}\) signal (\textit{bottom left}) with redshift, and the angular power spectrum of the 21 cm fluctuations at \(z=30\) (\textit{bottom right}), for 3 example DM decay channels with a decay lifetime $\tau = 1 \times 10^{26}$ s.}
    \label{fig:21 cm_Xe_Tb_T21_Cl_decay}%
\end{figure}
\par
As exemplified in Figure~\ref{fig:21 cm_Xe_Tb_T21_Cl_decay}, the amount of energy injection can vary greatly for a given lifetime, when the DM mass or decay channel is varied. For the three cases explored here, we find around \(\sim1\) order of magnitude variation in the forecasted relative errors for LRA, see Table~\ref{tab:decay_cases}. 
The forecasted errors at \(z=30\) can vary by almost 3 orders of magnitude between the most extreme cases. The results we found for \(f_{\rm{eff}}=0.1\) correspond roughly to 100 GeV decay to electron--positron pairs (\(\chi \rightarrow e^+e^-\)). However, for a 100 GeV particle decaying to photons only, the projected sensitivity is significantly weaker. Yet even in this case with minimal energy injection, LRA3 could reach almost percent-level precision constraints. On the other hand, there are decay scenarios which can inject much more energy than what our fiducial \(f_{\rm{eff}}=0.1\) describes e.g. 100 MeV \(\chi \rightarrow e^+e^-\), and so 21 cm LIM will be even more powerful to constrain certain cases. Therefore, our choice remains as a conservative assumption, not too optimistic nor pessimistic, given the different predictions across the available parameter space. Nevertheless, the main conclusions still hold: observations of the 21 cm power spectrum at \(z=30\) alone are unlikely to yield a \(S/N\) of unity for DM decay, except in decay cases where the energy injection is near maximal. In most cases, and especially to probe the longest decay lifetimes, reaching percent or sub-percent precision will require a lunar radio array with the ability to perform tomography during the cosmic dark ages. 21 cm LIM tomography with LRA will have the power to improve upon CMB constraints by several orders of magnitude - and for some decay channels, by even more than shown in our main results. The fact that the results are sensitive to the precise energy injection history demonstrates that the 21 cm power spectrum could be a powerful tool not only to detect a decay signal, but also constrain the DM particle properties, provided that potential degeneracies can be accounted for. A more detailed analysis, computing the \(f(z)\) functions over the whole redshift range (using e.g. \texttt{DarkHistory}~\cite{Liu:2019bbm} or \texttt{ExoCLASS}~\cite{Stocker:2018avm}) will be necessary once 21 cm data is realised in order to determine constraints on specific decay channels and masses.
 \par
In the case of annihilating DM, we do not assume a value for the efficiency factor $f_{\rm{eff}}$, as all the model-dependence of the energy injection rate is condensed into the annihilation efficiency parameter $p_{\rm{ann}}$. While it has been shown that a constant energy injection rate can capture with high precision the effect of DM annihilation on the CMB~\cite{Hutsi:2011vx, Galli:2011rz,Giesen:2012rp,Madhavacheril:2013cna,Slatyer:2015jla}, whether this approach is sufficient to describe the impact on the 21 cm fluctuations during the dark ages needs to be evaluated. Given the proof-of-concept nature of this work, we expect the uncertainties introduced by this assumption to be sub-dominant, but the validity of this approach should be revisited with a calculation of the redshift-dependent $f(z)$ (or $p_{\rm{ann}}(z)$) curves once precision 21 cm measurements are made possible. This is especially important at low redshifts when including the effects of structure formation where the energy deposition curves can change substantially~\cite{Slatyer:2012yq}.

\begin{table}[h!]
\centering
\vspace{5mm}
\ra{1.3}
\begin{tabular}{rc|ccc}
\hline
\multicolumn{1}{l}{Experiment} & Redshift range        & \multicolumn{3}{c}{Relative Error $ \sigma_{\tau}/ \tau: \quad \tau=1 \times 10^{26} $ s}                                                       \\
\multicolumn{1}{l}{}           & \multicolumn{1}{l|}{} & 100 MeV $\chi \rightarrow e^+e^-$ & 100 GeV $\chi \rightarrow e^+e^-$ & 100 GeV $\chi \rightarrow \gamma \gamma$ \\ \hline
SKA                            & $z=30$                & $1.14 \times 10^{2}$              & $4.10 \times 10^{3}$              & $1.04 \times 10^{5}$                     \\ \hline
aSKA                       & $z=30$                & $1.15 \times 10^{-2}$             & $2.42 \times 10^{-1}$             & $2.90$                                   \\ \hline
\multirow{2}{*}{LRA1}          & $z=30$                & $1.90 \times 10^{-1}$             & $6.56$                            & $1.70 \times 10^{2}$                     \\
                               & $30 < z < 200 $       & $6.19 \times 10^{-2}$             & $6.07 \times 10^{-1}$             & $5.06$                                   \\ \hline
\multirow{2}{*}{LRA2}          & $z=30$                & $1.05 \times 10^{-2}$             & $1.59 \times 10^{-1}$             & $1.23$                                   \\
                               & $30 < z < 200 $       & $4.23 \times 10^{-3}$             & $8.28 \times 10^{-3}$             & $7.15 \times 10^{-2}$                    \\ \hline
\multirow{2}{*}{LRA3}          & $z=30$                & $1.01 \times 10^{-2}$             & $1.05 \times 10^{-1}$             & $3.67 \times 10^{-1}$                    \\
                               & $30 < z < 200 $       & $1.60 \times 10^{-3}$             & $2.12 \times 10^{-3}$             & $1.80 \times 10^{-2}$                    \\ \hline
\end{tabular}\caption{Forecasted 68\% C.L. relative errors for 3 example decay channels with a lifetime $\tau=1 \times 10^{26}$  s for each experimental set-up measuring the 21 cm angular power spectrum.}
\label{tab:decay_cases}
\end{table}

\bibliography{refs21 cm}

\providecommand{\href}[2]{#2}\begingroup\raggedright\begin{thebibliography}{100}

\bibitem{Bergstrom:2000pn}
L.~Bergstrm, ``{Nonbaryonic dark matter: Observational evidence and detection
  methods},'' \href{http://dx.doi.org/10.1088/0034-4885/63/5/2r3}{{\em Rept.
  Prog. Phys.} {\bfseries 63} (2000) 793},
\href{http://arxiv.org/abs/hep-ph/0002126}{{\ttfamily arXiv:hep-ph/0002126
  [hep-ph]}}.

\bibitem{Bertone_2005}
G.~Bertone, D.~Hooper, and J.~Silk, ``Particle dark matter: evidence,
  candidates and constraints,''
  \href{http://dx.doi.org/10.1016/j.physrep.2004.08.031}{{\em Physics Reports}
  {\bfseries 405} no.~5-6, (Jan, 2005) 279--390}.
  \url{http://dx.doi.org/10.1016/j.physrep.2004.08.031}.

\bibitem{Ali-Haimoud:2019khd}
A.~Kashlinsky {\em et~al.}, ``{Electromagnetic probes of primordial black holes
  as dark matter},''
\href{http://arxiv.org/abs/1903.04424}{{\ttfamily arXiv:1903.04424
  [astro-ph.CO]}}.

\bibitem{Gluscevic:2019yal}
V.~Gluscevic {\em et~al.}, ``{Cosmological Probes of Dark Matter Interactions:
  The Next Decade},''
\href{http://arxiv.org/abs/1903.05140}{{\ttfamily arXiv:1903.05140
  [astro-ph.CO]}}.

\bibitem{Abazajian:2012ys}
K.~N. Abazajian {\em et~al.}, ``{Light Sterile Neutrinos: A White Paper},''
\href{http://arxiv.org/abs/1204.5379}{{\ttfamily arXiv:1204.5379 [hep-ph]}}.

\bibitem{Boyarsky:2018tvu}
A.~Boyarsky, M.~Drewes, T.~Lasserre, S.~Mertens, and O.~Ruchayskiy, ``{Sterile
  Neutrino Dark Matter},''
  \href{http://dx.doi.org/10.1016/j.ppnp.2018.07.004}{{\em Prog. Part. Nucl.
  Phys.} {\bfseries 104} (2019) 1--45},
\href{http://arxiv.org/abs/1807.07938}{{\ttfamily arXiv:1807.07938 [hep-ph]}}.

\bibitem{KIM200218}
H.~B. Kim and J.~E. Kim, ``Late decaying axino as CDM and its lifetime bound,''
  \href{http://dx.doi.org/https://doi.org/10.1016/S0370-2693(01)01507-6}{{\em
  Physics Letters B} {\bfseries 527} no.~1, (2002) 18 -- 22}.
  \url{http://www.sciencedirect.com/science/article/pii/S0370269301015076}.

\bibitem{Hamaguchi:2017ihw}
K.~Hamaguchi, K.~Nakayama, and Y.~Tang, ``{Gravitino/Axino as Decaying Dark
  Matter and Cosmological Tensions},''
  \href{http://dx.doi.org/10.1016/j.physletb.2017.06.071}{{\em Phys. Lett.}
  {\bfseries B772} (2017) 415--419},
\href{http://arxiv.org/abs/1705.04521}{{\ttfamily arXiv:1705.04521 [hep-ph]}}.

\bibitem{BEREZINSKY1991382}
V.~Berezinsky, A.~Masiero, and J.~Valle, ``Cosmological signatures of
  supersymmetry with spontaneously broken R parity,''
  \href{http://dx.doi.org/https://doi.org/10.1016/0370-2693(91)91055-Z}{{\em
  Physics Letters B} {\bfseries 266} no.~3, (1991) 382 -- 388}.
  \url{http://www.sciencedirect.com/science/article/pii/037026939191055Z}.

\bibitem{Finkbeiner:2007kk}
D.~P. Finkbeiner and N.~Weiner, ``{Exciting Dark Matter and the INTEGRAL/SPI
  511 keV signal},'' \href{http://dx.doi.org/10.1103/PhysRevD.76.083519}{{\em
  Phys. Rev.} {\bfseries D76} (2007) 083519},
\href{http://arxiv.org/abs/astro-ph/0702587}{{\ttfamily arXiv:astro-ph/0702587
  [astro-ph]}}.

\bibitem{Finkbeiner:2014sja}
D.~P. Finkbeiner and N.~Weiner, ``{X-ray line from exciting dark matter},''
  \href{http://dx.doi.org/10.1103/PhysRevD.94.083002}{{\em Phys. Rev.}
  {\bfseries D94} no.~8, (2016) 083002},
\href{http://arxiv.org/abs/1402.6671}{{\ttfamily arXiv:1402.6671 [hep-ph]}}.

\bibitem{Kaplan:2009de}
D.~E. Kaplan, G.~Z. Krnjaic, K.~R. Rehermann, and C.~M. Wells, ``{Atomic Dark
  Matter},'' \href{http://dx.doi.org/10.1088/1475-7516/2010/05/021}{{\em JCAP}
  {\bfseries 1005} (2010) 021},
\href{http://arxiv.org/abs/0909.0753}{{\ttfamily arXiv:0909.0753 [hep-ph]}}.

\bibitem{Jungman:1995df}
G.~Jungman, M.~Kamionkowski, and K.~Griest, ``{Supersymmetric dark matter},''
  \href{http://dx.doi.org/10.1016/0370-1573(95)00058-5}{{\em Phys. Rept.}
  {\bfseries 267} (1996) 195--373},
\href{http://arxiv.org/abs/hep-ph/9506380}{{\ttfamily arXiv:hep-ph/9506380
  [hep-ph]}}.

\bibitem{Gaskins_2016}
J.~M. Gaskins, ``A review of indirect searches for particle dark matter,''
  \href{http://dx.doi.org/10.1080/00107514.2016.1175160}{{\em Contemporary
  Physics} {\bfseries 57} no.~4, (Jun, 2016) 496--525}.
  \url{http://dx.doi.org/10.1080/00107514.2016.1175160}.

\bibitem{Schumann:2019eaa}
M.~Schumann, ``{Direct Detection of WIMP Dark Matter: Concepts and Status},''
  \href{http://dx.doi.org/10.1088/1361-6471/ab2ea5}{{\em J. Phys.} {\bfseries
  G46} no.~10, (2019) 103003},
\href{http://arxiv.org/abs/1903.03026}{{\ttfamily arXiv:1903.03026
  [astro-ph.CO]}}.

\bibitem{Lin:2019uvt}
T.~Lin, ``{Dark matter models and direct detection},''
  \href{http://dx.doi.org/10.22323/1.333.0009}{{\em PoS} {\bfseries 333} (2019)
  009},
\href{http://arxiv.org/abs/1904.07915}{{\ttfamily arXiv:1904.07915 [hep-ph]}}.

\bibitem{Boveia:2018yeb}
A.~Boveia and C.~Doglioni, ``{Dark Matter Searches at Colliders},''
  \href{http://dx.doi.org/10.1146/annurev-nucl-101917-021008}{{\em Ann. Rev.
  Nucl. Part. Sci.} {\bfseries 68} (2018) 429--459},
\href{http://arxiv.org/abs/1810.12238}{{\ttfamily arXiv:1810.12238 [hep-ex]}}.

\bibitem{Slatyer:2017sev}
T.~R. {Slatyer}, ``{TASI Lectures on Indirect Detection of Dark Matter},'' {\em
  arXiv e-prints} (Oct, 2017) arXiv:1710.05137,
  \href{http://arxiv.org/abs/1710.05137}{{\ttfamily arXiv:1710.05137
  [hep-ph]}}.

\bibitem{Galli:2009zc}
S.~Galli, F.~Iocco, G.~Bertone, and A.~Melchiorri, ``{CMB constraints on Dark
  Matter models with large annihilation cross-section},''
  \href{http://dx.doi.org/10.1103/PhysRevD.80.023505}{{\em Phys. Rev.}
  {\bfseries D80} (2009) 023505},
\href{http://arxiv.org/abs/0905.0003}{{\ttfamily arXiv:0905.0003
  [astro-ph.CO]}}.

\bibitem{Slatyer:2009yq}
T.~R. Slatyer, N.~Padmanabhan, and D.~P. Finkbeiner, ``{CMB Constraints on WIMP
  Annihilation: Energy Absorption During the Recombination Epoch},''
  \href{http://dx.doi.org/10.1103/PhysRevD.80.043526}{{\em Phys. Rev.}
  {\bfseries D80} (2009) 043526},
\href{http://arxiv.org/abs/0906.1197}{{\ttfamily arXiv:0906.1197
  [astro-ph.CO]}}.

\bibitem{Huetsi:2009ex}
G.~Huetsi, A.~Hektor, and M.~Raidal, ``{Constraints on leptonically
  annihilating Dark Matter from reionization and extragalactic gamma
  background},'' \href{http://dx.doi.org/10.1051/0004-6361/200912760}{{\em
  Astron. Astrophys.} {\bfseries 505} (2009) 999--1005},
\href{http://arxiv.org/abs/0906.4550}{{\ttfamily arXiv:0906.4550
  [astro-ph.CO]}}.

\bibitem{Galli:2011rz}
S.~Galli, F.~Iocco, G.~Bertone, and A.~Melchiorri, ``{Updated CMB constraints
  on Dark Matter annihilation cross-sections},''
  \href{http://dx.doi.org/10.1103/PhysRevD.84.027302}{{\em Phys. Rev.}
  {\bfseries D84} (2011) 027302},
\href{http://arxiv.org/abs/1106.1528}{{\ttfamily arXiv:1106.1528
  [astro-ph.CO]}}.

\bibitem{Finkbeiner:2011dx}
D.~P. Finkbeiner, S.~Galli, T.~Lin, and T.~R. Slatyer, ``{Searching for Dark
  Matter in the CMB: A Compact Parameterization of Energy Injection from New
  Physics},'' \href{http://dx.doi.org/10.1103/PhysRevD.85.043522}{{\em Phys.
  Rev.} {\bfseries D85} (2012) 043522},
\href{http://arxiv.org/abs/1109.6322}{{\ttfamily arXiv:1109.6322
  [astro-ph.CO]}}.

\bibitem{Ade:2015xua}
{\bfseries Planck} Collaboration, P.~A.~R. Ade {\em et~al.}, ``{Planck 2015
  results. XIII. Cosmological parameters},''
  \href{http://dx.doi.org/10.1051/0004-6361/201525830}{{\em Astron. Astrophys.}
  {\bfseries 594} (2016) A13},
\href{http://arxiv.org/abs/1502.01589}{{\ttfamily arXiv:1502.01589
  [astro-ph.CO]}}.

\bibitem{Slatyer:2015jla}
T.~R. Slatyer, ``{Indirect dark matter signatures in the cosmic dark ages. I.
  Generalizing the bound on s-wave dark matter annihilation from Planck
  results},'' \href{http://dx.doi.org/10.1103/PhysRevD.93.023527}{{\em Phys.
  Rev.} {\bfseries D93} no.~2, (2016) 023527},
\href{http://arxiv.org/abs/1506.03811}{{\ttfamily arXiv:1506.03811 [hep-ph]}}.

\bibitem{Poulin:2016anj}
V.~Poulin, J.~Lesgourgues, and P.~D. Serpico, ``{Cosmological constraints on
  exotic injection of electromagnetic energy},''
  \href{http://dx.doi.org/10.1088/1475-7516/2017/03/043}{{\em JCAP} {\bfseries
  1703} no.~03, (2017) 043},
\href{http://arxiv.org/abs/1610.10051}{{\ttfamily arXiv:1610.10051
  [astro-ph.CO]}}.

\bibitem{Aghanim:2018eyx}
{\bfseries Planck} Collaboration, N.~Aghanim {\em et~al.}, ``{Planck 2018
  results. VI. Cosmological parameters},''
\href{http://arxiv.org/abs/1807.06209}{{\ttfamily arXiv:1807.06209
  [astro-ph.CO]}}.

\bibitem{Ellis:1984eq}
J.~R. Ellis, J.~E. Kim, and D.~V. Nanopoulos, ``{Cosmological Gravitino
  Regeneration and Decay},''
\href{http://dx.doi.org/10.1016/0370-2693(84)90334-4}{{\em Phys. Lett.}
  {\bfseries 145B} (1984) 181--186}.

\bibitem{Sarkar:1984tt}
S.~Sarkar and A.~M. Cooper-Sarkar, ``{Cosmological and experimental constraints
  on the tau neutrino},''
  \href{http://dx.doi.org/10.1016/0370-2693(84)90101-1}{{\em Phys. Lett.}
  {\bfseries 148B} (1984) 347--354}.
[,I.362(1984)].

\bibitem{Hu:1993gc}
W.~Hu and J.~Silk, ``{Thermalization constraints and spectral distortions for
  massive unstable relic particles},''
\href{http://dx.doi.org/10.1103/PhysRevLett.70.2661}{{\em Phys. Rev. Lett.}
  {\bfseries 70} (1993) 2661--2664}.

\bibitem{McDonald:2000bk}
P.~McDonald, R.~J. Scherrer, and T.~P. Walker, ``{Cosmic microwave background
  constraint on residual annihilations of relic particles},''
  \href{http://dx.doi.org/10.1103/PhysRevD.63.023001}{{\em Phys. Rev.}
  {\bfseries D63} (2001) 023001},
\href{http://arxiv.org/abs/astro-ph/0008134}{{\ttfamily arXiv:astro-ph/0008134
  [astro-ph]}}.

\bibitem{Chluba:2011hw}
J.~Chluba and R.~A. Sunyaev, ``{The evolution of CMB spectral distortions in
  the early Universe},''
  \href{http://dx.doi.org/10.1111/j.1365-2966.2011.19786.x}{{\em Mon. Not. Roy.
  Astron. Soc.} {\bfseries 419} (2012) 1294--1314},
\href{http://arxiv.org/abs/1109.6552}{{\ttfamily arXiv:1109.6552
  [astro-ph.CO]}}.

\bibitem{Chluba:2013wsa}
J.~Chluba, ``{Distinguishing different scenarios of early energy release with
  spectral distortions of the cosmic microwave background},''
  \href{http://dx.doi.org/10.1093/mnras/stt1733}{{\em Mon. Not. Roy. Astron.
  Soc.} {\bfseries 436} (2013) 2232--2243},
\href{http://arxiv.org/abs/1304.6121}{{\ttfamily arXiv:1304.6121
  [astro-ph.CO]}}.

\bibitem{Chluba:2013pya}
J.~Chluba and D.~Jeong, ``{Teasing bits of information out of the CMB energy
  spectrum},'' \href{http://dx.doi.org/10.1093/mnras/stt2327}{{\em Mon. Not.
  Roy. Astron. Soc.} {\bfseries 438} no.~3, (2014) 2065--2082},
\href{http://arxiv.org/abs/1306.5751}{{\ttfamily arXiv:1306.5751
  [astro-ph.CO]}}.

\bibitem{Iocco:2008va}
F.~Iocco, G.~Mangano, G.~Miele, O.~Pisanti, and P.~D. Serpico, ``{Primordial
  Nucleosynthesis: from precision cosmology to fundamental physics},''
  \href{http://dx.doi.org/10.1016/j.physrep.2009.02.002}{{\em Phys. Rept.}
  {\bfseries 472} (2009) 1--76},
\href{http://arxiv.org/abs/0809.0631}{{\ttfamily arXiv:0809.0631 [astro-ph]}}.

\bibitem{Pospelov:2010hj}
M.~Pospelov and J.~Pradler, ``{Big Bang Nucleosynthesis as a Probe of New
  Physics},'' \href{http://dx.doi.org/10.1146/annurev.nucl.012809.104521}{{\em
  Ann. Rev. Nucl. Part. Sci.} {\bfseries 60} (2010) 539--568},
\href{http://arxiv.org/abs/1011.1054}{{\ttfamily arXiv:1011.1054 [hep-ph]}}.

\bibitem{Poulin:2015opa}
V.~Poulin and P.~D. Serpico, ``{Nonuniversal BBN bounds on electromagnetically
  decaying particles},''
  \href{http://dx.doi.org/10.1103/PhysRevD.91.103007}{{\em Phys. Rev.}
  {\bfseries D91} no.~10, (2015) 103007},
\href{http://arxiv.org/abs/1503.04852}{{\ttfamily arXiv:1503.04852
  [astro-ph.CO]}}.

\bibitem{Slatyer:2016qyl}
T.~R. Slatyer and C.-L. Wu, ``{General Constraints on Dark Matter Decay from
  the Cosmic Microwave Background},''
  \href{http://dx.doi.org/10.1103/PhysRevD.95.023010}{{\em Phys. Rev.}
  {\bfseries D95} no.~2, (2017) 023010},
\href{http://arxiv.org/abs/1610.06933}{{\ttfamily arXiv:1610.06933
  [astro-ph.CO]}}.

\bibitem{Lucca:2019rxf}
M.~Lucca, N.~Schneberg, D.~C. Hooper, J.~Lesgourgues, and J.~Chluba, ``{The
  synergy between CMB spectral distortions and anisotropies},''
\href{http://arxiv.org/abs/1910.04619}{{\ttfamily arXiv:1910.04619
  [astro-ph.CO]}}.

\bibitem{Cohen:2016uyg}
T.~Cohen, K.~Murase, N.~L. Rodd, B.~R. Safdi, and Y.~Soreq, ``{Gamma-ray
  Constraints on Decaying Dark Matter and Implications for IceCube},''
  \href{http://dx.doi.org/10.1103/PhysRevLett.119.021102}{{\em Phys. Rev.
  Lett.} {\bfseries 119} no.~2, (2017) 021102},
\href{http://arxiv.org/abs/1612.05638}{{\ttfamily arXiv:1612.05638 [hep-ph]}}.

\bibitem{Blanco:2018esa}
C.~Blanco and D.~Hooper, ``{Constraints on Decaying Dark Matter from the
  Isotropic Gamma-Ray Background},''
  \href{http://dx.doi.org/10.1088/1475-7516/2019/03/019}{{\em JCAP} {\bfseries
  1903} no.~03, (2019) 019},
\href{http://arxiv.org/abs/1811.05988}{{\ttfamily arXiv:1811.05988
  [astro-ph.HE]}}.

\bibitem{Giesen:2012rp}
G.~Giesen, J.~Lesgourgues, B.~Audren, and Y.~Ali-Haimoud, ``{CMB photons
  shedding light on dark matter},''
  \href{http://dx.doi.org/10.1088/1475-7516/2012/12/008}{{\em JCAP} {\bfseries
  1212} (2012) 008},
\href{http://arxiv.org/abs/1209.0247}{{\ttfamily arXiv:1209.0247
  [astro-ph.CO]}}.

\bibitem{Lopez-Honorez:2013lcm}
L.~Lopez-Honorez, O.~Mena, S.~Palomares-Ruiz, and A.~C. Vincent, ``{Constraints
  on dark matter annihilation from CMB observations before Planck},''
  \href{http://dx.doi.org/10.1088/1475-7516/2013/07/046}{{\em JCAP} {\bfseries
  1307} (2013) 046},
\href{http://arxiv.org/abs/1303.5094}{{\ttfamily arXiv:1303.5094
  [astro-ph.CO]}}.

\bibitem{Kawasaki:2015peu}
M.~Kawasaki, K.~Nakayama, and T.~Sekiguchi, ``{CMB Constraint on Dark Matter
  Annihilation after Planck 2015},''
  \href{http://dx.doi.org/10.1016/j.physletb.2016.03.005}{{\em Phys. Lett.}
  {\bfseries B756} (2016) 212--215},
\href{http://arxiv.org/abs/1512.08015}{{\ttfamily arXiv:1512.08015
  [astro-ph.CO]}}.

\bibitem{Abramowski:2014tra}
{\bfseries H.E.S.S.} Collaboration, A.~Abramowski {\em et~al.}, ``{Search for
  dark matter annihilation signatures in H.E.S.S. observations of Dwarf
  Spheroidal Galaxies},''
  \href{http://dx.doi.org/10.1103/PhysRevD.90.112012}{{\em Phys. Rev.}
  {\bfseries D90} (2014) 112012},
\href{http://arxiv.org/abs/1410.2589}{{\ttfamily arXiv:1410.2589
  [astro-ph.HE]}}.

\bibitem{Ackermann:2015zua}
{\bfseries Fermi-LAT} Collaboration, M.~Ackermann {\em et~al.}, ``{Searching
  for Dark Matter Annihilation from Milky Way Dwarf Spheroidal Galaxies with
  Six Years of Fermi Large Area Telescope Data},''
  \href{http://dx.doi.org/10.1103/PhysRevLett.115.231301}{{\em Phys. Rev.
  Lett.} {\bfseries 115} no.~23, (2015) 231301},
\href{http://arxiv.org/abs/1503.02641}{{\ttfamily arXiv:1503.02641
  [astro-ph.HE]}}.

\bibitem{Ahnen:2016qkx}
{\bfseries MAGIC, Fermi-LAT} Collaboration, M.~L. Ahnen {\em et~al.}, ``{Limits
  to Dark Matter Annihilation Cross-Section from a Combined Analysis of MAGIC
  and Fermi-LAT Observations of Dwarf Satellite Galaxies},''
  \href{http://dx.doi.org/10.1088/1475-7516/2016/02/039}{{\em JCAP} {\bfseries
  1602} no.~02, (2016) 039},
\href{http://arxiv.org/abs/1601.06590}{{\ttfamily arXiv:1601.06590
  [astro-ph.HE]}}.

\bibitem{Fermi-LAT:2016uux}
{\bfseries Fermi-LAT, DES} Collaboration, A.~Albert {\em et~al.}, ``{Searching
  for Dark Matter Annihilation in Recently Discovered Milky Way Satellites with
  Fermi-LAT},'' \href{http://dx.doi.org/10.3847/1538-4357/834/2/110}{{\em
  Astrophys. J.} {\bfseries 834} no.~2, (2017) 110},
\href{http://arxiv.org/abs/1611.03184}{{\ttfamily arXiv:1611.03184
  [astro-ph.HE]}}.

\bibitem{Archambault:2017wyh}
{\bfseries VERITAS} Collaboration, S.~Archambault {\em et~al.}, ``{Dark Matter
  Constraints from a Joint Analysis of Dwarf Spheroidal Galaxy Observations
  with VERITAS},'' \href{http://dx.doi.org/10.1103/PhysRevD.95.082001}{{\em
  Phys. Rev.} {\bfseries D95} no.~8, (2017) 082001},
\href{http://arxiv.org/abs/1703.04937}{{\ttfamily arXiv:1703.04937
  [astro-ph.HE]}}.

\bibitem{Leane:2018kjk}
R.~K. Leane, T.~R. Slatyer, J.~F. Beacom, and K.~C.~Y. Ng, ``{GeV-scale thermal
  WIMPs: Not even slightly ruled out},''
  \href{http://dx.doi.org/10.1103/PhysRevD.98.023016}{{\em Phys. Rev.}
  {\bfseries D98} no.~2, (2018) 023016},
\href{http://arxiv.org/abs/1805.10305}{{\ttfamily arXiv:1805.10305 [hep-ph]}}.

\bibitem{Madhavacheril:2013cna}
M.~S. Madhavacheril, N.~Sehgal, and T.~R. Slatyer, ``{Current Dark Matter
  Annihilation Constraints from CMB and Low-Redshift Data},''
  \href{http://dx.doi.org/10.1103/PhysRevD.89.103508}{{\em Phys. Rev.}
  {\bfseries D89} (2014) 103508},
\href{http://arxiv.org/abs/1310.3815}{{\ttfamily arXiv:1310.3815
  [astro-ph.CO]}}.

\bibitem{Green:2018pmd}
D.~Green, P.~D. Meerburg, and J.~Meyers, ``{Aspects of Dark Matter Annihilation
  in Cosmology},'' \href{http://dx.doi.org/10.1088/1475-7516/2019/04/025}{{\em
  JCAP} {\bfseries 1904} (2019) 025},
\href{http://arxiv.org/abs/1804.01055}{{\ttfamily arXiv:1804.01055
  [astro-ph.CO]}}.

\bibitem{Kovetz:2017agg}
E.~D. Kovetz {\em et~al.}, ``{Line-Intensity Mapping: 2017 Status Report},''
\href{http://arxiv.org/abs/1709.09066}{{\ttfamily arXiv:1709.09066
  [astro-ph.CO]}}.

\bibitem{Kovetz:2019uss}
E.~D. Kovetz {\em et~al.}, ``{Astrophysics and Cosmology with Line-Intensity
  Mapping},''
\href{http://arxiv.org/abs/1903.04496}{{\ttfamily arXiv:1903.04496
  [astro-ph.CO]}}.

\bibitem{Bernal:2019jdo}
J.~L. Bernal, P.~C. Breysse, H.~Gil-Marn, and E.~D. Kovetz, ``{A User's Guide
  to Extracting Cosmological Information from Line-Intensity Maps},''
\href{http://arxiv.org/abs/1907.10067}{{\ttfamily arXiv:1907.10067
  [astro-ph.CO]}}.

\bibitem{Bernal:2019gfq}
J.~L. Bernal, P.~C. Breysse, and E.~D. Kovetz, ``{The Cosmic Expansion History
  from Line-Intensity Mapping},''
\href{http://arxiv.org/abs/1907.10065}{{\ttfamily arXiv:1907.10065
  [astro-ph.CO]}}.

\bibitem{Munoz:2019fkt}
J.~B. Muoz, ``{Standard Ruler at Cosmic Dawn},''
  \href{http://dx.doi.org/10.1103/PhysRevLett.123.131301}{{\em Phys. Rev.
  Lett.} {\bfseries 123} no.~13, (2019) 131301},
\href{http://arxiv.org/abs/1904.07868}{{\ttfamily arXiv:1904.07868
  [astro-ph.CO]}}.

\bibitem{Furlanetto:2006jb}
S.~Furlanetto, S.~P. Oh, and F.~Briggs, ``{Cosmology at Low Frequencies: The 21
  cm Transition and the High-Redshift Universe},''
  \href{http://dx.doi.org/10.1016/j.physrep.2006.08.002}{{\em Phys. Rept.}
  {\bfseries 433} (2006) 181--301},
\href{http://arxiv.org/abs/astro-ph/0608032}{{\ttfamily arXiv:astro-ph/0608032
  [astro-ph]}}.

\bibitem{Pritchard:2011xb}
J.~R. Pritchard and A.~Loeb, ``{21-cm cosmology},''
  \href{http://dx.doi.org/10.1088/0034-4885/75/8/086901}{{\em Rept. Prog.
  Phys.} {\bfseries 75} (2012) 086901},
\href{http://arxiv.org/abs/1109.6012}{{\ttfamily arXiv:1109.6012
  [astro-ph.CO]}}.

\bibitem{Breysse:2018slj}
P.~C. Breysse, Y.~Ali-Hamoud, and C.~M. Hirata, ``{Ultimate frontier of 21-cm
  cosmology},'' \href{http://dx.doi.org/10.1103/PhysRevD.98.043520}{{\em Phys.
  Rev.} {\bfseries D98} no.~4, (2018) 043520},
\href{http://arxiv.org/abs/1804.10626}{{\ttfamily arXiv:1804.10626
  [astro-ph.CO]}}.

\bibitem{Furlanetto:2006wp}
S.~R. Furlanetto, S.~P. Oh, and E.~Pierpaoli, ``{The Effects of Dark Matter
  Decay and Annihilation on the High-Redshift 21 cm Background},''
  \href{http://dx.doi.org/10.1103/PhysRevD.74.103502}{{\em Phys. Rev.}
  {\bfseries D74} (2006) 103502},
\href{http://arxiv.org/abs/astro-ph/0608385}{{\ttfamily arXiv:astro-ph/0608385
  [astro-ph]}}.

\bibitem{Shchekinov:2006eb}
Y.~A. Shchekinov and E.~O. Vasiliev, ``{Particle decay in the early universe:
  predictions for 21 cm},''
  \href{http://dx.doi.org/10.1111/j.1365-2966.2007.11715.x}{{\em Mon. Not. Roy.
  Astron. Soc.} {\bfseries 379} (2007) 1003--1010},
\href{http://arxiv.org/abs/astro-ph/0604231}{{\ttfamily arXiv:astro-ph/0604231
  [astro-ph]}}.

\bibitem{Valdes:2007cu}
M.~Valdes, A.~Ferrara, M.~Mapelli, and E.~Ripamonti, ``{Constraining DM through
  21 cm observations},''
  \href{http://dx.doi.org/10.1111/j.1365-2966.2007.11594.x}{{\em Mon. Not. Roy.
  Astron. Soc.} {\bfseries 377} (2007) 245--252},
\href{http://arxiv.org/abs/astro-ph/0701301}{{\ttfamily arXiv:astro-ph/0701301
  [astro-ph]}}.

\bibitem{Valdes:2012zv}
M.~Valdes, C.~Evoli, A.~Mesinger, A.~Ferrara, and N.~Yoshida, ``{The nature of
  dark matter from the global high redshift HI 21 cm signal},''
  \href{http://dx.doi.org/10.1093/mnras/sts458}{{\em Mon. Not. Roy. Astron.
  Soc.} {\bfseries 429} (2013) 1705--1716},
\href{http://arxiv.org/abs/1209.2120}{{\ttfamily arXiv:1209.2120
  [astro-ph.CO]}}.

\bibitem{Ali-Haimoud:2013hpa}
Y.~Ali-Haimoud, P.~D. Meerburg, and S.~Yuan, ``{New light on 21 cm intensity
  fluctuations from the dark ages},''
  \href{http://dx.doi.org/10.1103/PhysRevD.89.083506}{{\em Phys. Rev.}
  {\bfseries D89} no.~8, (2014) 083506},
\href{http://arxiv.org/abs/1312.4948}{{\ttfamily arXiv:1312.4948
  [astro-ph.CO]}}.

\bibitem{Evoli:2014pva}
C.~Evoli, A.~Mesinger, and A.~Ferrara, ``{Unveiling the nature of dark matter
  with high redshift 21 cm line experiments},''
  \href{http://dx.doi.org/10.1088/1475-7516/2014/11/024}{{\em JCAP} {\bfseries
  1411} no.~11, (2014) 024},
\href{http://arxiv.org/abs/1408.1109}{{\ttfamily arXiv:1408.1109
  [astro-ph.HE]}}.

\bibitem{Lopez-Honorez:2016sur}
L.~Lopez-Honorez, O.~Mena, n.~Molin, S.~Palomares-Ruiz, and A.~C. Vincent,
  ``{The 21 cm signal and the interplay between dark matter annihilations and
  astrophysical processes},''
  \href{http://dx.doi.org/10.1088/1475-7516/2016/08/004}{{\em JCAP} {\bfseries
  1608} no.~08, (2016) 004},
\href{http://arxiv.org/abs/1603.06795}{{\ttfamily arXiv:1603.06795
  [astro-ph.CO]}}.

\bibitem{Poulin:2016nat}
V.~Poulin, P.~D. Serpico, and J.~Lesgourgues, ``{A fresh look at linear
  cosmological constraints on a decaying dark matter component},''
  \href{http://dx.doi.org/10.1088/1475-7516/2016/08/036}{{\em JCAP} {\bfseries
  1608} no.~08, (2016) 036},
\href{http://arxiv.org/abs/1606.02073}{{\ttfamily arXiv:1606.02073
  [astro-ph.CO]}}.

\bibitem{Cirelli:2009bb}
M.~Cirelli, F.~Iocco, and P.~Panci, ``{Constraints on Dark Matter annihilations
  from reionization and heating of the intergalactic gas},''
  \href{http://dx.doi.org/10.1088/1475-7516/2009/10/009}{{\em JCAP} {\bfseries
  0910} (2009) 009},
\href{http://arxiv.org/abs/0907.0719}{{\ttfamily arXiv:0907.0719
  [astro-ph.CO]}}.

\bibitem{Liu:2016cnk}
H.~Liu, T.~R. Slatyer, and J.~Zavala, ``{Contributions to cosmic reionization
  from dark matter annihilation and decay},''
  \href{http://dx.doi.org/10.1103/PhysRevD.94.063507}{{\em Phys. Rev.}
  {\bfseries D94} no.~6, (2016) 063507},
\href{http://arxiv.org/abs/1604.02457}{{\ttfamily arXiv:1604.02457
  [astro-ph.CO]}}.

\bibitem{Cohen:2016jbh}
A.~Cohen, A.~Fialkov, R.~Barkana, and M.~Lotem, ``{Charting the Parameter Space
  of the Global 21-cm Signal},''
  \href{http://dx.doi.org/10.1093/mnras/stx2065}{{\em Mon. Not. Roy. Astron.
  Soc.} {\bfseries 472} no.~2, (2017) 1915--1931},
\href{http://arxiv.org/abs/1609.02312}{{\ttfamily arXiv:1609.02312
  [astro-ph.CO]}}.

\bibitem{Cohen:2017xpx}
A.~Cohen, A.~Fialkov, and R.~Barkana, ``{Charting the Parameter Space of the
  21-cm Power Spectrum},'' \href{http://dx.doi.org/10.1093/mnras/sty1094}{{\em
  Mon. Not. Roy. Astron. Soc.} {\bfseries 478} no.~2, (2018) 2193--2217},
\href{http://arxiv.org/abs/1709.02122}{{\ttfamily arXiv:1709.02122
  [astro-ph.CO]}}.

\bibitem{Burns:2019zia}
J.~O. Burns {\em et~al.}, ``{Dark Cosmology: Investigating Dark Matter \&
  Exotic Physics in the Dark Ages using the Redshifted 21-cm Global
  Spectrum},''
\href{http://arxiv.org/abs/1902.06147}{{\ttfamily arXiv:1902.06147
  [astro-ph.CO]}}.

\bibitem{Furlanetto:2019jso}
S.~Furlanetto {\em et~al.}, ``{Astro 2020 Science White Paper: Fundamental
  Cosmology in the Dark Ages with 21-cm Line Fluctuations},''
\href{http://arxiv.org/abs/1903.06212}{{\ttfamily arXiv:1903.06212
  [astro-ph.CO]}}.

\bibitem{Munoz:2016owz}
J.~B. Muoz, E.~D. Kovetz, A.~Raccanelli, M.~Kamionkowski, and J.~Silk,
  ``{Towards a measurement of the spectral runnings},''
  \href{http://dx.doi.org/10.1088/1475-7516/2017/05/032}{{\em JCAP} {\bfseries
  1705} (2017) 032},
\href{http://arxiv.org/abs/1611.05883}{{\ttfamily arXiv:1611.05883
  [astro-ph.CO]}}.

\bibitem{Pourtsidou:2016ctq}
A.~Pourtsidou, ``{Synergistic tests of inflation},''
\href{http://arxiv.org/abs/1612.05138}{{\ttfamily arXiv:1612.05138
  [astro-ph.CO]}}.

\bibitem{Sekiguchi:2017cdy}
T.~Sekiguchi, T.~Takahashi, H.~Tashiro, and S.~Yokoyama, ``{21 cm Angular Power
  Spectrum from Minihalos as a Probe of Primordial Spectral Runnings},''
  \href{http://dx.doi.org/10.1088/1475-7516/2018/02/053}{{\em JCAP} {\bfseries
  1802} no.~02, (2018) 053},
\href{http://arxiv.org/abs/1705.00405}{{\ttfamily arXiv:1705.00405
  [astro-ph.CO]}}.

\bibitem{Munoz:2015eqa}
J.~B. Muoz, Y.~Ali-Hamoud, and M.~Kamionkowski, ``{Primordial non-gaussianity
  from the bispectrum of 21-cm fluctuations in the dark ages},''
  \href{http://dx.doi.org/10.1103/PhysRevD.92.083508}{{\em Phys. Rev.}
  {\bfseries D92} no.~8, (2015) 083508},
\href{http://arxiv.org/abs/1506.04152}{{\ttfamily arXiv:1506.04152
  [astro-ph.CO]}}.

\bibitem{Bernal:2017nec}
J.~L. Bernal, A.~Raccanelli, L.~Verde, and J.~Silk, ``{Signatures of primordial
  black holes as seeds of supermassive black holes},''
  \href{http://dx.doi.org/10.1088/1475-7516/2018/05/017}{{\em JCAP} {\bfseries
  1805} no.~05, (2018) 017},
\href{http://arxiv.org/abs/1712.01311}{{\ttfamily arXiv:1712.01311
  [astro-ph.CO]}}.

\bibitem{Mena:2019nhm}
O.~Mena, S.~Palomares-Ruiz, P.~Villanueva-Domingo, and S.~J. Witte,
  ``{Constraining the primordial black hole abundance with 21-cm cosmology},''
  \href{http://dx.doi.org/10.1103/PhysRevD.100.043540}{{\em Phys. Rev.}
  {\bfseries D100} no.~4, (2019) 043540},
\href{http://arxiv.org/abs/1906.07735}{{\ttfamily arXiv:1906.07735
  [astro-ph.CO]}}.

\bibitem{Bacon:2018dui}
{\bfseries SKA} Collaboration, D.~J. Bacon {\em et~al.}, ``{Cosmology with
  Phase 1 of the Square Kilometre Array: Red Book 2018: Technical
  specifications and performance forecasts},'' {\em Submitted to: Publ. Astron.
  Soc. Austral.} (2018) ,
\href{http://arxiv.org/abs/1811.02743}{{\ttfamily arXiv:1811.02743
  [astro-ph.CO]}}.

\bibitem{Carilli:2007eb}
C.~L. Carilli, J.~N. Hewitt, and A.~Loeb, ``{Low frequency radio astronomy from
  the moon: Cosmic reionization and more},'' in {\em {Workshop on Astrophysics
  Enabled by the Return to the Moon Baltimore, Maryland, November 28-30,
  2006}}.
\newblock 2007.
\newblock
\href{http://arxiv.org/abs/astro-ph/0702070}{{\ttfamily arXiv:astro-ph/0702070
  [ASTRO-PH]}}.
\newblock

\bibitem{Jester_2009}
S.~Jester and H.~Falcke, ``Science with a lunar low-frequency array: From the
  dark ages of the Universe to nearby exoplanets,''
  \href{http://dx.doi.org/10.1016/j.newar.2009.02.001}{{\em New Astronomy
  Reviews} {\bfseries 53} no.~1-2, (May, 2009) 1--26}.
  \url{http://dx.doi.org/10.1016/j.newar.2009.02.001}.

\bibitem{Burns:2012bv}
J.~O. Burns, T.~J.~W. Lazio, and W.~Bottke, ``{Astrophysics Conducted by the
  Lunar University Network for Astrophysics Research (LUNAR) and the Center for
  Lunar Origins (CLOE)},''
\href{http://arxiv.org/abs/1209.2233}{{\ttfamily arXiv:1209.2233
  [astro-ph.CO]}}.

\bibitem{LRA_whitepaper}
J.~Silk, A.~Raccanelli, C.~S., M.~J., , and E.~D. Kovetz, ``A White Paper on
  Cosmology with a Lunar Radio Array,'' {\em in preparation} .

\bibitem{Wouthuysen_1952}
S.~A. Wouthuysen, ``On the excitation mechanism of the 21-cm (radio-frequency)
  interstellar hydrogen emission line.,''
  \href{http://dx.doi.org/10.1086/106661}{{\em The Astronomical Journal}
  {\bfseries 57} (1952) 31}. \url{http://dx.doi.org/10.1086/106661}.

\bibitem{Field_1959}
G.~B. Field, ``The Spin Temperature of Intergalactic Neutral Hydrogen.,''
  \href{http://dx.doi.org/10.1086/146653}{{\em The Astrophysical Journal}
  {\bfseries 129} (May, 1959) 536}. \url{http://dx.doi.org/10.1086/146653}.

\bibitem{Hirata:2005mz}
C.~M. Hirata, ``{Wouthuysen-Field coupling strength and application to
  high-redshift 21 cm radiation},''
  \href{http://dx.doi.org/10.1111/j.1365-2966.2005.09949.x}{{\em Mon. Not. Roy.
  Astron. Soc.} {\bfseries 367} (2006) 259--274},
\href{http://arxiv.org/abs/astro-ph/0507102}{{\ttfamily arXiv:astro-ph/0507102
  [astro-ph]}}.

\bibitem{Field_1958}
G.~Field, ``Excitation of the Hydrogen 21-CM Line,''
  \href{http://dx.doi.org/10.1109/jrproc.1958.286741}{{\em Proceedings of the
  IRE} {\bfseries 46} no.~1, (Jan, 1958) 240--250}.
  \url{http://dx.doi.org/10.1109/JRPROC.1958.286741}.

\bibitem{Kuhlen:2005cm}
M.~Kuhlen, P.~Madau, and R.~Montgomery, ``{The spin temperature and 21cm
  brightness of the intergalactic medium in the pre-reionization era},''
  \href{http://dx.doi.org/10.1086/500548}{{\em Astrophys. J.} {\bfseries 637}
  (2006) L1--L4},
\href{http://arxiv.org/abs/astro-ph/0510814}{{\ttfamily arXiv:astro-ph/0510814
  [astro-ph]}}.

\bibitem{Liszt:2001kh}
H.~Liszt, ``{The spin temperature of warm interstellar h I},''
  \href{http://dx.doi.org/10.1051/0004-6361:20010395}{{\em Astron. Astrophys.}
  {\bfseries 371} (2001) 698},
\href{http://arxiv.org/abs/astro-ph/0103246}{{\ttfamily arXiv:astro-ph/0103246
  [astro-ph]}}.

\bibitem{Poulin:2015pna}
V.~Poulin, P.~D. Serpico, and J.~Lesgourgues, ``{Dark Matter annihilations in
  halos and high-redshift sources of reionization of the universe},''
  \href{http://dx.doi.org/10.1088/1475-7516/2015/12/041}{{\em JCAP} {\bfseries
  1512} no.~12, (2015) 041},
\href{http://arxiv.org/abs/1508.01370}{{\ttfamily arXiv:1508.01370
  [astro-ph.CO]}}.

\bibitem{Pillepich:2006fj}
A.~Pillepich, C.~Porciani, and S.~Matarrese, ``{The bispectrum of redshifted
  21-cm fluctuations from the dark ages},''
  \href{http://dx.doi.org/10.1086/517963}{{\em Astrophys. J.} {\bfseries 662}
  (2007) 1--14},
\href{http://arxiv.org/abs/astro-ph/0611126}{{\ttfamily arXiv:astro-ph/0611126
  [astro-ph]}}.

\bibitem{Bharadwaj:2004nr}
S.~Bharadwaj and S.~S. Ali, ``{The CMBR fluctuations from HI perturbations
  prior to reionization},''
  \href{http://dx.doi.org/10.1111/j.1365-2966.2004.07907.x}{{\em Mon. Not. Roy.
  Astron. Soc.} {\bfseries 352} (2004) 142},
\href{http://arxiv.org/abs/astro-ph/0401206}{{\ttfamily arXiv:astro-ph/0401206
  [astro-ph]}}.

\bibitem{1953ApJ...117..134L}
D.~N. {Limber}, ``{The Analysis of Counts of the Extragalactic Nebulae in Terms
  of a Fluctuating Density Field.},''
  \href{http://dx.doi.org/10.1086/145672}{{\em The Astrophysical Journal}
  {\bfseries 117} (Jan., 1953) 134}.

\bibitem{LoVerde:2008re}
M.~LoVerde and N.~Afshordi, ``{Extended Limber Approximation},''
  \href{http://dx.doi.org/10.1103/PhysRevD.78.123506}{{\em Phys. Rev.}
  {\bfseries D78} (2008) 123506},
\href{http://arxiv.org/abs/0809.5112}{{\ttfamily arXiv:0809.5112 [astro-ph]}}.

\bibitem{Blas:2011rf}
D.~Blas, J.~Lesgourgues, and T.~Tram, ``{The Cosmic Linear Anisotropy Solving
  System (CLASS) II: Approximation schemes},''
  \href{http://dx.doi.org/10.1088/1475-7516/2011/07/034}{{\em JCAP} {\bfseries
  1107} (2011) 034},
\href{http://arxiv.org/abs/1104.2933}{{\ttfamily arXiv:1104.2933
  [astro-ph.CO]}}.

\bibitem{Chluba:2010ca}
J.~Chluba and R.~M. Thomas, ``{Towards a complete treatment of the cosmological
  recombination problem},''
  \href{http://dx.doi.org/10.1111/j.1365-2966.2010.17940.x}{{\em Mon. Not. Roy.
  Astron. Soc.} {\bfseries 412} (2011) 748},
\href{http://arxiv.org/abs/1010.3631}{{\ttfamily arXiv:1010.3631
  [astro-ph.CO]}}.

\bibitem{Peebles:1968ja}
P.~J.~E. Peebles, ``{Recombination of the Primeval Plasma},''
\href{http://dx.doi.org/10.1086/149628}{{\em Astrophys. J.} {\bfseries 153}
  (1968) 1}.

\bibitem{Seager:1999bc}
S.~Seager, D.~D. Sasselov, and D.~Scott, ``{A new calculation of the
  recombination epoch},'' \href{http://dx.doi.org/10.1086/312250}{{\em
  Astrophys. J.} {\bfseries 523} (1999) L1--L5},
\href{http://arxiv.org/abs/astro-ph/9909275}{{\ttfamily arXiv:astro-ph/9909275
  [astro-ph]}}.

\bibitem{Chluba:2009uv}
J.~Chluba, ``{Could the Cosmological Recombination Spectrum Help Us Understand
  Annihilating Dark Matter?},''
  \href{http://dx.doi.org/10.1111/j.1365-2966.2009.15957.x}{{\em Mon. Not. Roy.
  Astron. Soc.} {\bfseries 402} (2010) 1195},
\href{http://arxiv.org/abs/0910.3663}{{\ttfamily arXiv:0910.3663
  [astro-ph.CO]}}.

\bibitem{AliHaimoud:2010dx}
Y.~Ali-Haimoud and C.~M. Hirata, ``{HyRec: A fast and highly accurate
  primordial hydrogen and helium recombination code},''
  \href{http://dx.doi.org/10.1103/PhysRevD.83.043513}{{\em Phys. Rev.}
  {\bfseries D83} (2011) 043513},
\href{http://arxiv.org/abs/1011.3758}{{\ttfamily arXiv:1011.3758
  [astro-ph.CO]}}.

\bibitem{Slatyer:2012yq}
T.~R. Slatyer, ``{Energy Injection And Absorption In The Cosmic Dark Ages},''
  \href{http://dx.doi.org/10.1103/PhysRevD.87.123513}{{\em Phys. Rev.}
  {\bfseries D87} no.~12, (2013) 123513},
\href{http://arxiv.org/abs/1211.0283}{{\ttfamily arXiv:1211.0283
  [astro-ph.CO]}}.

\bibitem{Chen:2003gz}
X.-L. Chen and M.~Kamionkowski, ``{Particle decays during the cosmic dark
  ages},'' \href{http://dx.doi.org/10.1103/PhysRevD.70.043502}{{\em Phys. Rev.}
  {\bfseries D70} (2004) 043502},
\href{http://arxiv.org/abs/astro-ph/0310473}{{\ttfamily arXiv:astro-ph/0310473
  [astro-ph]}}.

\bibitem{Galli:2013dna}
S.~Galli, T.~R. Slatyer, M.~Valdes, and F.~Iocco, ``{Systematic Uncertainties
  In Constraining Dark Matter Annihilation From The Cosmic Microwave
  Background},'' \href{http://dx.doi.org/10.1103/PhysRevD.88.063502}{{\em Phys.
  Rev.} {\bfseries D88} (2013) 063502},
\href{http://arxiv.org/abs/1306.0563}{{\ttfamily arXiv:1306.0563
  [astro-ph.CO]}}.

\bibitem{Slatyer:2015kla}
T.~R. Slatyer, ``{Indirect Dark Matter Signatures in the Cosmic Dark Ages II.
  Ionization, Heating and Photon Production from Arbitrary Energy
  Injections},'' \href{http://dx.doi.org/10.1103/PhysRevD.93.023521}{{\em Phys.
  Rev.} {\bfseries D93} no.~2, (2016) 023521},
\href{http://arxiv.org/abs/1506.03812}{{\ttfamily arXiv:1506.03812
  [astro-ph.CO]}}.

\bibitem{Liu:2019bbm}
H.~Liu, G.~W. Ridgway, and T.~R. Slatyer, ``{DarkHistory: A code package for
  calculating modified cosmic ionization and thermal histories with dark matter
  and other exotic energy injections},''
\href{http://arxiv.org/abs/1904.09296}{{\ttfamily arXiv:1904.09296
  [astro-ph.CO]}}.

\bibitem{Hisano:2004ds}
J.~Hisano, S.~Matsumoto, M.~M. Nojiri, and O.~Saito, ``{Non-perturbative effect
  on dark matter annihilation and gamma ray signature from galactic center},''
  \href{http://dx.doi.org/10.1103/PhysRevD.71.063528}{{\em Phys. Rev.}
  {\bfseries D71} (2005) 063528},
\href{http://arxiv.org/abs/hep-ph/0412403}{{\ttfamily arXiv:hep-ph/0412403
  [hep-ph]}}.

\bibitem{10.2307/2342435}
R.~A. Fisher, ``The Logic of Inductive Inference,'' {\em Journal of the Royal
  Statistical Society} {\bfseries 98} no.~1, (1935) 39--82.
  \url{http://www.jstor.org/stable/2342435}.

\bibitem{Tegmark:1996bz}
M.~Tegmark, A.~Taylor, and A.~Heavens, ``{Karhunen-Loeve eigenvalue problems in
  cosmology: How should we tackle large data sets?},''
  \href{http://dx.doi.org/10.1086/303939}{{\em Astrophys. J.} {\bfseries 480}
  (1997) 22},
\href{http://arxiv.org/abs/astro-ph/9603021}{{\ttfamily arXiv:astro-ph/9603021
  [astro-ph]}}.

\bibitem{Jaffe:2000yt}
A.~H. Jaffe, M.~Kamionkowski, and L.-M. Wang, ``{Polarization pursuers'
  guide},'' \href{http://dx.doi.org/10.1103/PhysRevD.61.083501}{{\em Phys.
  Rev.} {\bfseries D61} (2000) 083501},
\href{http://arxiv.org/abs/astro-ph/9909281}{{\ttfamily arXiv:astro-ph/9909281
  [astro-ph]}}.

\bibitem{Knox:2002pe}
L.~Knox and Y.-S. Song, ``{A Limit on the detectability of the energy scale of
  inflation},'' \href{http://dx.doi.org/10.1103/PhysRevLett.89.011303}{{\em
  Phys. Rev. Lett.} {\bfseries 89} (2002) 011303},
\href{http://arxiv.org/abs/astro-ph/0202286}{{\ttfamily arXiv:astro-ph/0202286
  [astro-ph]}}.

\bibitem{Kesden:2002ku}
M.~Kesden, A.~Cooray, and M.~Kamionkowski, ``{Separation of gravitational wave
  and cosmic shear contributions to cosmic microwave background
  polarization},'' \href{http://dx.doi.org/10.1103/PhysRevLett.89.011304}{{\em
  Phys. Rev. Lett.} {\bfseries 89} (2002) 011304},
\href{http://arxiv.org/abs/astro-ph/0202434}{{\ttfamily arXiv:astro-ph/0202434
  [astro-ph]}}.

\bibitem{Zaldarriaga:2003du}
M.~Zaldarriaga, S.~R. Furlanetto, and L.~Hernquist, ``{21 Centimeter
  fluctuations from cosmic gas at high redshifts},''
  \href{http://dx.doi.org/10.1086/386327}{{\em Astrophys. J.} {\bfseries 608}
  (2004) 622--635},
\href{http://arxiv.org/abs/astro-ph/0311514}{{\ttfamily arXiv:astro-ph/0311514
  [astro-ph]}}.

\bibitem{2017MNRAS.464.4995M}
T.~J. {Mozdzen}, J.~D. {Bowman}, R.~A. {Monsalve}, and A.~E.~E. {Rogers},
  ``{Improved measurement of the spectral index of the diffuse radio background
  between 90 and 190 MHz},''
  \href{http://dx.doi.org/10.1093/mnras/stw2696}{{\em Mon. Not. Roy. Astron.
  Soc.} {\bfseries 464} no.~4, (Feb, 2017) 4995--5002},
  \href{http://arxiv.org/abs/1609.08705}{{\ttfamily arXiv:1609.08705
  [astro-ph.IM]}}.

\bibitem{Calore:2014nla}
F.~Calore, I.~Cholis, C.~McCabe, and C.~Weniger, ``{A Tale of Tails: Dark
  Matter Interpretations of the Fermi GeV Excess in Light of Background Model
  Systematics},'' \href{http://dx.doi.org/10.1103/PhysRevD.91.063003}{{\em
  Phys. Rev.} {\bfseries D91} no.~6, (2015) 063003},
\href{http://arxiv.org/abs/1411.4647}{{\ttfamily arXiv:1411.4647 [hep-ph]}}.

\bibitem{Cholis:2019ejx}
I.~Cholis, T.~Linden, and D.~Hooper, ``{A Robust Excess in the Cosmic-Ray
  Antiproton Spectrum: Implications for Annihilating Dark Matter},''
  \href{http://dx.doi.org/10.1103/PhysRevD.99.103026}{{\em Phys. Rev.}
  {\bfseries D99} no.~10, (2019) 103026},
\href{http://arxiv.org/abs/1903.02549}{{\ttfamily arXiv:1903.02549
  [astro-ph.HE]}}.

\bibitem{Abazajian:2016yjj}
{\bfseries CMB-S4} Collaboration, K.~N. Abazajian {\em et~al.}, ``{CMB-S4
  Science Book, First Edition},''
\href{http://arxiv.org/abs/1610.02743}{{\ttfamily arXiv:1610.02743
  [astro-ph.CO]}}.

\bibitem{Bulbul:2014sua}
E.~Bulbul, M.~Markevitch, A.~Foster, R.~K. Smith, M.~Loewenstein, and S.~W.
  Randall, ``{Detection of An Unidentified Emission Line in the Stacked X-ray
  spectrum of Galaxy Clusters},''
  \href{http://dx.doi.org/10.1088/0004-637X/789/1/13}{{\em Astrophys. J.}
  {\bfseries 789} (2014) 13},
\href{http://arxiv.org/abs/1402.2301}{{\ttfamily arXiv:1402.2301
  [astro-ph.CO]}}.

\bibitem{Boyarsky:2014jta}
A.~Boyarsky, O.~Ruchayskiy, D.~Iakubovskyi, and J.~Franse, ``{Unidentified Line
  in X-Ray Spectra of the Andromeda Galaxy and Perseus Galaxy Cluster},''
  \href{http://dx.doi.org/10.1103/PhysRevLett.113.251301}{{\em Phys. Rev.
  Lett.} {\bfseries 113} (2014) 251301},
\href{http://arxiv.org/abs/1402.4119}{{\ttfamily arXiv:1402.4119
  [astro-ph.CO]}}.

\bibitem{Kong:2014gea}
J.-C. Park, S.~C. Park, and K.~Kong, ``{X-ray line signal from 7 keV axino dark
  matter decay},'' \href{http://dx.doi.org/10.1016/j.physletb.2014.04.037}{{\em
  Phys. Lett.} {\bfseries B733} (2014) 217--220},
\href{http://arxiv.org/abs/1403.1536}{{\ttfamily arXiv:1403.1536 [hep-ph]}}.

\bibitem{Adhikari:2016bei}
M.~Drewes {\em et~al.}, ``{A White Paper on keV Sterile Neutrino Dark
  Matter},'' \href{http://dx.doi.org/10.1088/1475-7516/2017/01/025}{{\em JCAP}
  {\bfseries 1701} no.~01, (2017) 025},
\href{http://arxiv.org/abs/1602.04816}{{\ttfamily arXiv:1602.04816 [hep-ph]}}.

\bibitem{Cosme:2017cxk}
C.~Cosme, J.~G. Rosa, and O.~Bertolami, ``{Scalar field dark matter with
  spontaneous symmetry breaking and the $3.5$ keV line},''
  \href{http://dx.doi.org/10.1016/j.physletb.2018.04.062}{{\em Phys. Lett.}
  {\bfseries B781} (2018) 639--644},
\href{http://arxiv.org/abs/1709.09674}{{\ttfamily arXiv:1709.09674 [hep-ph]}}.

\bibitem{Stocker:2018avm}
P.~Stcker, M.~Krmer, J.~Lesgourgues, and V.~Poulin, ``{Exotic energy injection
  with ExoCLASS: Application to the Higgs portal model and evaporating black
  holes},'' \href{http://dx.doi.org/10.1088/1475-7516/2018/03/018}{{\em JCAP}
  {\bfseries 1803} no.~03, (2018) 018},
\href{http://arxiv.org/abs/1801.01871}{{\ttfamily arXiv:1801.01871
  [astro-ph.CO]}}.

\bibitem{Munoz:2015bca}
J.~B. Muoz, E.~D. Kovetz, and Y.~Ali-Hamoud, ``{Heating of Baryons due to
  Scattering with Dark Matter During the Dark Ages},''
  \href{http://dx.doi.org/10.1103/PhysRevD.92.083528}{{\em Phys. Rev.}
  {\bfseries D92} no.~8, (2015) 083528},
\href{http://arxiv.org/abs/1509.00029}{{\ttfamily arXiv:1509.00029
  [astro-ph.CO]}}.

\bibitem{Tashiro:2014tsa}
H.~Tashiro, K.~Kadota, and J.~Silk, ``{Effects of dark matter-baryon scattering
  on redshifted 21 cm signals},''
  \href{http://dx.doi.org/10.1103/PhysRevD.90.083522}{{\em Phys. Rev.}
  {\bfseries D90} no.~8, (2014) 083522},
\href{http://arxiv.org/abs/1408.2571}{{\ttfamily arXiv:1408.2571
  [astro-ph.CO]}}.

\bibitem{DeBoer:2016tnn}
D.~R. DeBoer {\em et~al.}, ``{Hydrogen Epoch of Reionization Array (HERA)},''
  \href{http://dx.doi.org/10.1088/1538-3873/129/974/045001}{{\em Publ. Astron.
  Soc. Pac.} {\bfseries 129} no.~974, (2017) 045001},
\href{http://arxiv.org/abs/1606.07473}{{\ttfamily arXiv:1606.07473
  [astro-ph.IM]}}.

\bibitem{Gardner:2006ky}
J.~P. Gardner {\em et~al.}, ``{The James Webb Space Telescope},''
  \href{http://dx.doi.org/10.1007/s11214-006-8315-7}{{\em Space Sci. Rev.}
  {\bfseries 123} (2006) 485},
\href{http://arxiv.org/abs/astro-ph/0606175}{{\ttfamily arXiv:astro-ph/0606175
  [astro-ph]}}.

\bibitem{Burns:2011wf}
J.~O. Burns, T.~J.~W. Lazio, S.~D. Bale, J.~D. Bowman, R.~F. Bradley, C.~L.
  Carilli, S.~R. Furlanetto, G.~J.~A. Harker, A.~Loeb, and J.~R. Pritchard,
  ``{Probing the First Stars and Black Holes in the Early Universe with the
  Dark Ages Radio Explorer (DARE)},''
  \href{http://dx.doi.org/10.1016/j.asr.2011.10.014}{{\em Adv. Space Res.}
  {\bfseries 49} (2012) 433--450},
\href{http://arxiv.org/abs/1106.5194}{{\ttfamily arXiv:1106.5194
  [astro-ph.CO]}}.

\bibitem{Chluba:2015lpa}
J.~Chluba, D.~Paoletti, F.~Finelli, and J.-A. Rubio-Martn, ``{Effect of
  primordial magnetic fields on the ionization history},''
  \href{http://dx.doi.org/10.1093/mnras/stv1096}{{\em Mon. Not. Roy. Astron.
  Soc.} {\bfseries 451} no.~2, (2015) 2244--2250},
\href{http://arxiv.org/abs/1503.04827}{{\ttfamily arXiv:1503.04827
  [astro-ph.CO]}}.

\bibitem{Hutsi:2011vx}
G.~Hutsi, J.~Chluba, A.~Hektor, and M.~Raidal, ``{WMAP7 and future CMB
  constraints on annihilating dark matter: implications on GeV-scale WIMPs},''
  \href{http://dx.doi.org/10.1051/0004-6361/201116914}{{\em Astron. Astrophys.}
  {\bfseries 535} (2011) A26},
\href{http://arxiv.org/abs/1103.2766}{{\ttfamily arXiv:1103.2766
  [astro-ph.CO]}}.

\end{thebibliography}\endgroup
\bibliographystyle{utcaps}

\end{document}